\documentclass{amsart}
\usepackage{amsmath, amsthm}
\usepackage{amsfonts,amssymb,amscd}

\newtheorem{theorem}{Theorem}[section]
\newtheorem{lemma}[theorem]{Lemma}
\newtheorem{proposition}[theorem]{Proposition}

\newtheorem{corollary}[theorem]{Corollary}

\newenvironment{bew}[2]{\removelastskip\vspace{6pt}\noindent
 {\it Proof  #1.}~\rm#2}{\par\vspace{6pt}}

\numberwithin{equation}{section}

\font\sevenrm=cmr7
\newcommand{\hp}{\hphantom}
\newcommand{\wQ}{\widetilde Q}
\newcommand{\rv}{\mbox{v}}

\newcommand{\rvseven}{\mbox{\sevenrm v}}
\newcommand{\rwseven}{\mbox{\sevenrm w}}
\newcommand{\rw}{\mbox{w}}
\newcommand{\rz}{\mbox{z}}
\newcommand{\tq}{\tilde q}
\newcommand{\tr}{\tilde r}
\newcommand{\e}{\epsilon}
\newcommand{\fg}{\mathfrak{g}}
\newcommand{\fh}{\mathfrak{h}}
\newcommand{\fG}{\mathfrak{G}}
\newcommand{\ader}[1]{a^\alpha_{i_1\cdots i_{#1}}}
\newcommand{\Phibar}{\overline\Phi}
\newcommand{\Pvar}[4]{\Phi^{#2}{}_{\mathbf #1_{#3}}^{\mathbf #1_{#4}'}}
\newcommand{\Pvarbar}[4]{\overline{\Phi}^{#2}{}_{\mathbf #1_{#3}'}^{\mathbf #1_{
#4}}}
\newcommand{\Pder}[3]{\partial_{\Phi^{#1}}{}^{#2}_{#3}}
\newcommand{\Pderbar}[3]{\overline{\partial}_{\Phi^{#1}}{}^{#2}_{#3}}
\newcommand{\hook}
{\mathbin{\raise1.5pt\hbox{\hbox{{\vbox{\hrule height.4pt 
width6pt depth0pt}}}\vrule height3pt width.4pt depth0pt}\,}}

\newcommand{\pr}{\operatorname{pr}}

\newcommand{\J}[2]{J^{#1} #2}

\begin{document}
\title[Symmetries of Yang-Mills fields]{
Classification of generalized symmetries of the Yang-Mills
fields with a semi-simple structure group
}
\author{Juha Pohjanpelto}
\address{Department of Mathematics\\
           Oregon State University\\
           Corvallis, Oregon 97331-4605\\
           USA}

\email{juha@math.orst.edu}
\begin{abstract}
A complete classification of
generalized symmetries of the Yang-Mills equations
on Minkowski space
with a semi-simple structure group is carried out.
It is shown that any generalized symmetry, up
to a generalized gauge symmetry, agrees with a
first order symmetry on solutions of the Yang-Mills
equations. Let $\fg=\fg_1+\cdots+\fg_n$ be the decomposition
of the Lie algebra $\fg$ of the structure group into simple
ideals. First order symmetries for $\fg$-valued Yang-Mills
fields are found to consist of gauge symmetries,
conformal symmetries for $\fg_m$-valued Yang-Mills
fields, $1\leq m\leq n$, and their images under a
complex structure of $\fg_m$.
\end{abstract}

\maketitle
\section{Introduction}
\label{intro}
Generalized, or Lie-B\"acklund, 
symmetries of a system of 
differential equations, roughly speaking, 
are infinitesimal transformations 
involving the independent and 
dependent variables and their derivatives 
that preserve solutions to
the system. Since their introduction by Emmy 
Noether in her study of the correspondence
between  transformations preserving the
fundamental integral of a variational principle
and conservation laws of the associated Euler-Lagrange
equations, generalized symmetries have become
increasingly important in the geometric analysis
of differential equations. Besides the original
application to the identification of conservation
laws, generalized symmetries also play an 
important role in the study of infinite 
dimensional Hamiltonian systems \cite{Olver86}
and in various methods, in particular
in separation of variables \cite{Kalnins86},
\cite{Miller77}, for constructing
explicit solutions to differential equations.
There also seems to be a close connection
between completely integrable equations 
and generalized symmetries. B\"acklund
transformations have been shown to give
rise to infinite sequences of generalized
symmetries \cite{Kumei75}
and the existence of infinite 
number of independent generalized symmetries has,
in fact, been proposed as a test for the 
complete integrability of differential equations 
\cite{Mikhailov91}.

In this paper we carry out a complete analysis
of generalized 
symmetries of the Yang-Mills
equations on Minkowski space with a semi-simple 
structure group. Due to its implications, some
of which are discussed below, classification of
generalized symmetries and conservation laws
of the Yang-Mills fields has  been listed by 
Tsujishita \cite{Tsujishita82} as a significant open 
problem in the formal geometry of differential 
equations.

Presently, surprisingly little seem to be known about
the existence of higher order symmetries and
conservation laws for the Yang-Mills equations.
By construction, the Yang-Mills equations 
admit symmetries arising from the conformal 
transformations of Minkowski space and 
from the gauge transformations of the potential. 
Schwartz \cite{Schwartz82} has used a computer algebra 
to verify that for the structure group $SU(2)$,
the conformal group 
together with gauge transformations forms the 
maximal group of Lie symmetries of the equations.
Under the Noether correspondence, the 
conformal transformations yield 15 independent
conservation laws for Yang-Mills fields \cite{Glassey79}, 
while, by Noether's second theorem, the gauge 
symmetries reflect a differential identity, 
a divergence identity, satisfied by the Yang-Mills 
equations.

A priori, one easily discovers evidence for the existence 
of hidden symmetries of the Yang-Mills equations. 
In a recent paper \cite{sAjP2}, it is found that the simplest 
case of the Yang-Mills fields, the free electromagnetic
field, possesses a family of new generalized 
symmetries, the representatives of which 
of order $r$, $r\geq2$,
correspond to Killing spinors of type $(r-2,r+2)$ 
on Minkowski space. 
Counterparts of these symmetries of order 2 
in a curved spacetime metric are discussed in 
\cite{Kalnins92}. Additional evidence is provided
by the self-dual Yang-Mills equations and its various 
completely integrable reductions, which admit 
an infinite number of independent hidden symmetries.
On the other hand, negative evidence for the existence
of additional generalized symmetries is provided
by \cite{Torre95}, in which natural symmetries, i.e., symmetries
that transform equivariantly under the Poincar\'e
and gauge groups, are analyzed. It is found that
the only natural symmetries are generalized gauge symmetries 
arising from natural functions. However, the naturality
requirement is quite restrictive and excludes, in
particular, the conformal symmetries of the Yang-Mills
equations.  

The foremost application of a symmetry 
analysis of the Yang-Mills equations is the identification 
of conservation laws for the equations by Noether's
theorem. However, a full classification of conservation
laws is complicated by the degeneracy of the equations, 
due to which the correspondence between 
equivalence classes of conservation laws 
and equivalence classes of characteristics for
conservation laws
fails to be one-to-one. However, a complete symmetry 
analysis will still be an integral step in the determination 
whether the Yang-Mills equations possess any 
additional non-trivial conservation laws
besides those provided by the conformal symmetries.
We plan to treat this issue in a future publication.

Vinogradov \cite{Vinogradov89} has argued that a nondegenerate 
system of nonlinear differential equations involving more 
than two independent variables do not admit generalized
symmetries. Of course, due to the divergence identity
satisfied by the Yang-Mills equations, Vinogradov's
arguments do not apply in the present problem. However,
the study of the symmetries of degenerate systems
may reveal whether under suitable assumptions 
Vinogradov's theorem admits a converse of some form. 


Spinorial methods are pivotal in our analysis of symmetries
of the Yang-Mills fields. Recently, spinor techniques have 
been employed in the symmetry analysis of the Einstein 
equations \cite{Anderson96}, in the analysis of natural symmetries
of the Yang-Mills equations \cite{Torre95} and of
symmetries and conservation laws of Maxwell's equations
in Minkowski space \cite{sAjP1}, \cite{sAjP2} and in curved
background metric \cite{Kalnins92}.  Spinor techniques are 
also applicable to the problem at hand. The main 
computational tool is the exact sets of fields of 
Penrose \cite{Penrose84}, which can be used as part of 
the coordinate system for the infinitely prolonged 
solution manifold determined by the Yang-Mills 
equations. According to Penrose, the symmetrized 
covariant derivatives of the Yang-Mills spinor uniquely 
determine all the unsymmetrized derivatives on the 
solution manifold, a considerable simplification over 
the corresponding tensor treatment. 

In this paper we will show that the conformal symmetries
and their variants provided by the complex structure
of a Lie algebra, when it exists, together with the
generalized gauge symmetries are, up to equivalence,
the only generalized symmetries admitted by the Yang-Mills
equations on Minkowski space with a semi-simple 
structure group.
Specifically, we will prove the following Theorem.

\begin{theorem}
\label{T:maintheorem}
Let $\fg$ be a semisimple
Lie algebra, and let $\fg = \fg_1+\cdots + \fg_n$ be
the decomposition of $\fg$ into simple ideals. Let 
$Q_i=
Q_i(x^j,a^\beta_{k},\dots,a^\beta_{k,l_1\cdots l_p})$ 
be a generalized symmetry of order $p$ 
for $\fg$-valued Yang-Mills fields
on Minkowski space. 
Then there is a first order symmetry 
$\widetilde Q_i = 
\widetilde Q_i(x^j, a^\beta_{k},a^\beta_{k,l})$ 
of the Yang-Mills equations and a $\fg$-valued 
function 
$\mathcal X=
\mathcal X(x^j,a^\beta_{k},\dots,a^\beta_{k,l_1\cdots l_{q}})$ 
so that

\[ Q_i = \widetilde Q_i +\nabla_i\mathcal X \]
on solutions of the Yang-Mills equations.
 
Moreover, the function $\mathcal X$
can be chosen so that the symmetry
$\widetilde Q_i$ 
is expressible as a sum
\[\widetilde Q_i = 
\widetilde Q_{1,i}+\cdots +\widetilde Q_{n,i}\] 
of first order symmetries $\widetilde Q_{m,i}$ for 
$\fg_m$-valued  Yang-Mills fields, where,
in the case $\fg_m$ is a real form of a
simple complex Lie algebra, the symmetry
\[\widetilde Q_{m,i} = \xi^j_m F_{m,ij}\] 
is a conformal symmetry, and, in the case 
$\fg_m$ is the realification of a simple complex 
Lie algebra with the complex structure 
$J_{m}$, the symmetry 
\[\widetilde Q_{m,i}=\xi^j_m F_{m,ij}+
\tau^j_m J_{m}F_{m,ij}\]
is a sum of a conformal symmetry and 
the image of a conformal symmetry under 
$J_{m}$. In the above
expressions 
$F_{m,ij}$ denotes the
field tensor for $\fg_m$-valued Yang-Mills
fields.
\end{theorem}

The Yang-Mills equations also possess the obvious
symmetries consisting of permutations of isomorphic
components in the decomposition of the Lie algebra
$\fg$ into simple ideals. These symmetries,
however, are discrete and do not satisfy the defining
equations for generalized symmetries, and, as such,
do not yield conservation laws under the Noether 
correspondence.

Our paper is organized as follows. In Section
\ref{S:prelim} we establish notation and give a summary of some
basic properties of generalized symmetries and
of semi-simple Lie algebras. We also introduce 
spinorial methods and derive several technical results 
needed in the sequel. Section \ref{S:proof} in its
entirety is dedicated to the proof of Theorem 
\ref{T:maintheorem}.
In the Appendix we give the proof of a derivative 
formula presented in Section \ref{S:prelim}. 

\section{Preliminaries}
\label{S:prelim}

In this Section
we establish notation and review some basic definitions
and results from the theory of symmetries of differential 
equations and of semi-simple Lie algebras most relevant 
for the topic of the paper at hand. We also introduce
spinorial methods and present several technical
results needed in the proof of Theorem
\ref{T:maintheorem}.
For more details, see, for example, 
\cite{Helgason78}, 
\cite{Olver86},
\cite{Penrose84}, \cite{Ward90}.

Let $M$ be Minkowski space with coordinates 
$x^i$, $i=0,1,2,3$, and let $\fG$
be a Lie group with Lie algebra $\fg$. Recall that $\fG$, 
$\fg$ are called simple if $\fg$ is not abelian and admits
no ideals besides $\{0\}$ and $\fg$, and that $\fG$, $\fg$
are called semi-simple if the Killing form 
$\kappa(\rv,\rw) = tr(\mbox{ad}\,\rv\,\circ\,\mbox{ad}\, \rw)$ of $\fg$ is
non-degenerate, where $\mbox{ad}\, \rv(\mbox{z}) = [\rv,\mbox{z}]$.
A semi-simple Lie algebra $\fg$ is a direct sum
of simple ideals of $\fg$. We write
$A=\Lambda^1(M)\otimes \fg\to M$ 
for the bundle of $\fg$-valued Yang-Mills
potentials over $M$. 
Fix a basis $\{e_\alpha\}$ for $\fg$ and let $a^\alpha_i$
denote the components of the Yang-Mills potential. 
Then, as a coordinate bundle, 
$A=\{(x^i,a^\alpha_j)\}\to\{(x^i)\}$. We denote
the $p$th order jet bundle of local section of $A$
by $J^p(A)$, $0\leq p\leq\infty$. As a coordinate 
space, $J^p(A)$ is given by
\begin{equation}
\label{E:coordinates}
J^p(A) = \{(x^i,a^\alpha_i,a^\alpha_{i,j_1},
a^\alpha_{i,j_1j_2},\dots,a^\alpha_{i,j_1j_2\cdots j_p})\},
\end{equation}
where $a^\alpha_{i,j_1\cdots j_q}$ stands for the 
$q$th order derivative variables.

Write $\partial_i$ for the partial derivative 
operator $\partial/\partial x^i$ and define 
partial derivative operators 
$\partial_{a^\alpha}^{i,j_1\cdots j_p}$ by
\begin{eqnarray*}
&&\partial_{a^\alpha}^{i,j_1\cdots j_p}x^k = 0,\qquad\mbox{and}\\
&&\partial_{a^\alpha}^{i,j_1\cdots j_p}
a^\beta_{k,l_1\cdots l_q} =
\begin{cases}
\delta^{\beta}_\alpha\delta^i_k
\delta^{(j_1}_{l_1}\cdots\delta^{j_p)}_{l_p},
&\text{if $p=q$,}\\
0,&\text{if $p\neq q$.}\end{cases}
\end{eqnarray*}

A generalized vector field $X$ on $A$ is a vector field 
$$
X = P^i\partial_i+Q^{\alpha}_i\partial_{a^\alpha}^i,
$$
where the coefficients $P^i$, $Q^{\alpha}_i$ are 
differential functions on $A$, that is, functions 
on some $J^p(A)$, $p<\infty$. We call $p$ the
order of $X$.
An evolutionary vector field $Y$ is a generalized
vector field of the form
$$
Y = Q^\alpha_{i}\partial_{a^\alpha}^i.
$$
Here and in what follows we employ the standard 
Einstein summation convention.

Write $D_i$ for the total derivative operator
$$
D_i = \partial_i + 
\sum_{p\geq0}a^\alpha_{j,k_1\cdots k_pi}
\partial_{a^\alpha}^{j,k_1\cdots k_p}.
$$
The infinite prolongation $\pr X$ of $X$ 
is the unique lift of $X$ to a vector field
on $\J A$ preserving the contact ideal on 
$\J A$. In the coordinates \eqref{E:coordinates}, 
$\pr X$ is given by
$$
\pr X = P^iD_i+\sum_{p\geq 0}\bigl(D_{j_1}\cdots 
D_{j_p}Q^\alpha_{ev,i}\bigr)
\partial_{a^\alpha}^{i,j_1\cdots j_p},
$$
where the differential functions 
$Q^\alpha_{ev,i}$ are the  
components of the evolutionary form
$$
X_{ev} = (Q^\alpha_{i}-P^j a^\alpha_{i,j})
\partial_{a^\alpha}^i
$$
of $X$.

We extend the usual Yang-Mills covariant 
derivative to $\fg$-valued differential functions 
$G^\alpha$ by
$$
\nabla_iG^\alpha = D_iG^\alpha + [a_i,G]^\alpha.
$$
Then due to the Jacobi identity,
\begin{equation}
\label{E:covarbracket}
\nabla_i[G,H] = [\nabla_iG,H]+[G,\nabla_iH].
\end{equation}

We will use the following result repeatedly
in the proof of Theorem \ref{T:maintheorem}.
\begin{proposition} 
\label{P:productrule}
Let $G^\alpha$ be a $\fg$-valued and let
$r^\alpha_\beta$ be an {\rm End}($\fg$)-valued 
differential function.
Then
\begin{equation}
\label{E:productrule}
\nabla_i(r^\alpha_\beta G^\beta) = 
(D_i r^\alpha_\beta)G^\beta +
r^\alpha_\beta\nabla_i G^\beta-
(c^\alpha_{\beta\gamma}r^\beta_\delta+
r^\alpha_\beta c^\beta_{\gamma\delta})
a^\gamma_iG^\delta.
\end{equation}
In particular, if
\begin{equation}
\label{E:adjointcommute}
c^\alpha_{\beta\gamma}r^\beta_\delta+
r^\alpha_\beta c^\beta_{\gamma\delta}=0,
\end{equation}
then
\begin{equation}
\label{E:productrulesimple}
\nabla_i(r^\alpha_\beta G^\beta) = 
(D_i r^\alpha_\beta)G^\beta +
r^\alpha_\beta\nabla_i G^\beta.
\end{equation}
\end{proposition}

In order to classify first order symmetries of 
the Yang-Mills equations we need to analyze
equation \eqref{E:adjointcommute} more closely. 
Let $R:\fg\to \fg$ be the endomorphism of $\fg$ 
with the matrix $(r^\alpha_\beta)$. Then condition 
\eqref{E:adjointcommute} simply means that $R$ 
commutes with the adjoint representation, 
\begin{equation}
\label{E:Radjointcommute}
R\circ \mbox{ad}\,\rv = \mbox{ad}\,\rv\circ R
\qquad\mbox{for all }\rv\in \fg.
\end{equation}
If $\fg$ is a
simple complex Lie algebra, then Schur's
Lemma states that $R$ must be a scalar
multiple of the identity transformation.
However, the Yang-Mills fields in this
paper are real-valued, and we are thus lead
to consider the counterpart of Schur's 
Lemma for real Lie algebras.

Let $\fh$ be a complex Lie algebra. Then 
the realification $\fh^{\mathbf R}$ of $\fh$ 
is $\fh$ regarded as a real Lie algebra. 
A real form of $\fh$ is a subalgebra 
$\fh_o\subset \fh^{\mathbf R}$ so that
the complexification $\fh_o^{\mathbf C}$
is isomorphic with $\fh$, that is,
$\fh=\fh_o+\mbox{i}\fh_o$.

As is well known \cite{Helgason78}, a simple 
real algebra $\fg$ is either a real form $\fh_o$ 
or the realification $\fh^{\mathbf R}$
of a simple complex Lie algebra $\fh$.
In the latter case 
the complex 
structure of $\fh^{\mathbf R}$ is easily 
seen to commute with the adjoint
representation. As we will see,
for simple real Lie algebras the 
identity mapping and the complex 
structure are essentially the only 
endomorphisms commuting with the
adjoint representation.

Let
\begin{equation}
\label{E:gdecomp}
\fg = \fg_1+\cdots +\fg_n
\end{equation} 
be the decomposition of a semi-simple Lie algebra
$\fg$ into simple ideals $\fg_m$.
Write $P_m : \fg\to \fg_m$ for the projection
induced by the decomposition \eqref{E:gdecomp}.

\begin{proposition}
\label{P:adjointcommute}
Let 
$\fg$ be a semi-simple Lie algebra
as in \eqref{E:gdecomp}, and let $R$ be an 
endomorphism of $\fg$ commuting 
with the adjoint representation of $\fg$.
Then each $\fg_m$ is invariant under
$R$. Write $R_m$ for the restriction
of $R$ to $\fg_m$ so that
$R = R_1\circ P_1 + \cdots + R_n\circ P_n$. 
If $\fg_m$ is a real
form of a simple complex Lie algebra,
then $R_m = a_m\,\mbox{id}$ for some
$a_m\in\mathbf R$.
If $\fg_m$ is the realification of a simple
complex Lie algebra, then  
$R_m = a_m\,\mbox{id}+b_m\,J_m$ for some
$a_m$, $b_m\in\mathbf R$, where $J_m$ is
the complex structure of $\fg_m$.

Conversely, let $R_m :\fg_m\to \fg_m$ be as above.
Then the mapping 
$R = R_1\circ P_1 + \cdots + R_n\circ P_n$
commutes with the adjoint representation
of $\fg$.
\end{proposition}

\begin{proof} It is easy to see that the
image and the inverse image of an 
ideal in $\fg$ under $R$ are again 
ideals of $\fg$. Thus $R(\fg_m)$ is an
ideal of $\fg$, which we can assume to
be non-trivial. Hence $R(\fg_m)$ is a direct
sum of the members of a subfamily of
$\fg_1$, \dots, $\fg_n$. Write 
$R(\fg_m) = \fg_{p_1}+\cdots+ \fg_{p_q}$,
and let $R_m$ stand for the restriction of
$R$ to $\fg_m$.
If $R_m^{-1}(\fg_{p_r})\subset \fg_m$ is non-trivial,
then, since $\fg_m$ is simple, 
$R_m^{-1}(\fg_{p_r})=\fg_m$. Thus 
$R_m(\fg_m) = \fg_p$ for some $p$.
If $m\neq p$, then, by \eqref{E:Radjointcommute},
 we have that
$$
R_m([\rv,\rw]) = [\rv,R_m(\rw)] = 0\qquad
\mbox{for all }\rv,\rw\in \fg_m.
$$
But this contradicts the simplicity of
$\fg_m$ and the non-triviality of $R_m$. 
Thus $R(\fg_m) = \fg_m$.

Next suppose that $\fg_m$ is the real form 
$\fh_{m,o}$ of a complex simple Lie algebra $\fh_m$. 
Lift $R_m$ to the complexification $\fh_m$ of $\fg_m$ 
to obtain an endomorphism $R_m'$ of $\fh_m$
commuting with the adjoint representation of
$\fh_m$. Since $\fh_m$ is simple, we can use Schur's
lemma to conclude that there is $a_m\in\mathbf C$
such that $R_m'=a_m\,\mbox{id}$. But $R_m'$ must preserve
$\fg_m\subset \fh_m$, which implies that $a_m$ is real. 
Thus $R_m=a_m\,\mbox{id}$, and Schur's lemma
holds in this case.

Suppose in turn that $\fg_m$ is the realification
of a simple complex Lie algebra $\fh_m$. 
For simplicity, choose a basis 
$e_\alpha$, $1\leq \alpha\leq d_m$,
for example a Weyl basis, for $\fh_m$, in 
which the structure constants 
$c^\alpha_{\beta\gamma}$ are real.
Then, in the basis $e_\alpha$, 
$f_\alpha=\mbox{i}e_\alpha$, $1\leq \alpha\leq d_m$,
for $\fg_m$, the bracket relations are
\begin{equation}
\label{E:bracketJbasis}
[e_\alpha,e_\beta] = c_{\alpha\beta}^\gamma e_\gamma,\quad
[e_\alpha,f_\beta] = c_{\alpha\beta}^\gamma f_\gamma,\quad
[f_\alpha,f_\beta] = -c_{\alpha\beta}^\gamma e_\gamma.
\end{equation}
Let $t_1$, $t_2$ be the subspaces of 
$\fg_m^{\mathbf C}$ spanned by the vectors 
$$
k_\alpha = \frac{1}{2}(e_\alpha+\mbox{i}f_\alpha),\qquad
l_\alpha = \frac{1}{2}(e_\alpha-\mbox{i}f_\alpha),
\qquad\alpha=1,\dots, d_m,
$$
respectively. Then by \eqref{E:bracketJbasis},
$$
[k_\alpha,k_\beta] = 
c^\gamma_{\alpha\beta}k_\gamma,\qquad
[k_\alpha,l_\alpha] = 0,\qquad
[l_\alpha,l_\beta] = 
c^\gamma_{\alpha\beta}l_\gamma.
$$
Hence $t_1$, $t_2$ are isomorphic to $\fh_m$, 
and $\fg_m^{\mathbf C}$ is the direct sum of the 
ideals $t_1$, $t_2$.

As above, we show that $t_1$, $t_2$ 
are invariant under the lift $R'_m$ of
$R_m$ to $\fg_m^{\mathbf C}$. 
Now apply Schur's Lemma to $R'_m{}_{|t_k}$ to
see that there are $c_1$, $c_2\in\mathbf C$ so that 
$$
R_m' = c_1\, P_{t_1}+c_2\, P_{t_2},
$$
where $P_{t_k}:\fg_m^{\mathbf C}\to t_k$ 
is the projection induced by the direct sum
$\fg_m^{\mathbf C}=t_1+t_2$. 

We still need
to choose $c_1$, $c_2$ so that $R_m'$ 
preserves $\fg_m\subset \fg_m^{\mathbf C}$. Write
$$
c_1 = a_1+\mbox{i} b_1,\qquad
c_2 = a_2+\mbox{i} b_2.
$$
Then
$$
R_m'e_\alpha = 
\frac{1}{2}(a_1+a_2+\mbox{i}b_1+\mbox{i}b_2)e_\alpha +
\frac{1}{2}(-b_1+b_2+\mbox{i}a_1-\mbox{i}a_2)f_\alpha,
$$
which is contained in $\fg_m$ provided that 
$a_2=a_1$ and $b_2=-b_1$.
With this,
$$
R_m'f_\alpha = b_1e_\alpha+a_1f_\alpha.
$$
Thus, when $\fg_m$ is the realification of 
a simple complex Lie algebra, the space of 
endomorphisms of $\fg_m$ commuting with the
adjoint representation is spanned by the identity
transformation and the complex structure $J_m$
of $\fg_m$ given in the basis \eqref{E:bracketJbasis} by
$$
J_me_\alpha = f_\alpha,\qquad
J_mf_\alpha = -e_\alpha.
$$

The proof of the converse is now obvious.
This completes the proof of the Proposition.
\end{proof}

Write
\[
F^\alpha_{ij} = a^\alpha_{j,i}-a^\alpha_{i,j} + 
c^\alpha_{\beta\gamma}a^\beta_i a^\gamma_j
\]
for the components of the Yang-Mills field tensor.
The field tensor $F^\alpha_{ij}$ measures
the extent to which covariant derivatives
fail to commute. Specifically,
\begin{equation}
\label{E:covarcommute}
\nabla_i\nabla_jG-\nabla_j\nabla_iG = [F_{ij},G].
\end{equation}
The Yang-Mills equations and the Bianchi 
identity for $F_{ij}$ are
\begin{equation}
\label{E:YMequations}
\nabla^j F_{ij}=0\qquad\nabla^j*F_{ij}=0.
\end{equation}
Here and in what follows, we raise and 
lower indices using the Minkowski metric 
$\eta = \mbox{diag }(-1,1,1,1)$ and the symbol 
$*$ stands for the Hodge duality operator.
The Yang-Mills equations determine a submanifold
$\mathcal R\subset J^2(A)$, which we call the solution 
manifold of the equations. The $r$-fold, $r\leq p$, 
covariant derivatives
$$
\nabla_{i_1}\cdots\nabla_{i_r}\nabla^jF_{i_{r+1}j}=0
$$
of the field equations, in turn, determine the 
$p$-fold prolonged solution manifold 
$\mathcal R^p\subset J^{p+2}(A)$ of the equations.

A generalized symmetry of the Yang-Mills
equations of order $p$ is a generalized vector 
field $X$ of order $p$ satisfying 
\begin{equation}
\label{E:gensymm}
\pr X(\nabla^j F^\alpha_{ij}) = 0
\quad\mbox{on $\mathcal R^p$}.
\end{equation}
Note that any total vector field
\[P = P^iD_i = \pr \big(P^i\partial_i+
(P^ja_{i,j}^\alpha)\partial_{a^\alpha}^i\big),\] 
where $P^i$ are some differential functions,
satisfies the symmetry equations \eqref{E:gensymm}. 
Thus, in
particular, a generalized vector field $X$ is
a symmetry if and only if its evolutionary
form $X_{ev}$ is one. Hence
we only need to consider symmetries in 
evolutionary form. Equation \eqref{E:gensymm}, 
when written out for the components
of an evolutionary vector field 
$Q = Q^\alpha_i\partial_{a^\alpha}^i$
of order $p$, become
\begin{equation}
\label{E:evolsymmeq}
\nabla^j\nabla_iQ_j-
\nabla^j\nabla_jQ_i-
[F_i{}^j,Q_j] = 0
\quad\mbox{on $\mathcal R^p$}.
\end{equation}
We call this equation the determining equations
for symmetries of the Yang-Mills equations.
As is easily verified, the Yang-Mills equations 
admit generalized gauge symmetries given in 
component form by
\[
Q^\alpha_i = \nabla_i\mathcal X^\alpha,
\]
where $\mathcal X^\alpha$ is any $\fg$-valued 
differential function. Consequently,
we will call two generalized symmetries of the Yang-Mills
equations in evolutionary form equivalent if their difference 
agrees with a generalized gauge symmetry on some prolonged
solution manifold $\mathcal R^q$, $q\geq0$. 

By construction, the Yang-Mills equations also admit
symmetries arising from the conformal transformations
of the underlying Minkowski space. The components of 
the evolutionary form of the symmetry corresponding
to a conformal Killing vector $\xi^j$ is simply given by
\[
Q^\alpha_i = \xi^jF^\alpha_{ij}.
\]
If the decomposition of the semi-simple Lie algebra 
$\fg$ into a direct sum of simple ideals contains factors 
that are the realifications of simple complex Lie algebras, 
we can use the result of Proposition \ref{P:adjointcommute} 
to see that the 
Yang-Mills equations possess additional first order symmetries 
arising from the complex structures of these factors. The 
construction of the additional symmetries is based on the 
following result.

\begin{proposition}
\label{P:confsymm1}
Let $\fg$ be a Lie algebra
with structure constants $c^\alpha_{\beta\gamma}$
in some basis $e_\alpha$. Suppose that an
$\mbox{\rm End}(\fg)$-valued vector field 
$\xi^\alpha_\beta{}^j$ on Minkowski space
satisfies the equations
\[
\partial^{(i}\xi^{|\alpha|}_\beta{}^{j)} =
\eta^{ij}k^\alpha_\beta,\qquad
\xi^\alpha_\beta{}^j c^\beta_{\gamma\delta}+
c^\alpha_{\beta\gamma}\xi^\beta_{\delta}{}^j=0
\]
for some functions $k^\alpha_\beta$.
Then 
\begin{equation}
\label{E:confsymm1}
Q^\alpha_i = \xi^\alpha_\beta{}^jF^\beta_{ij}
\end{equation}
are the components of a first order generalized
symmetry of the Yang-Mills equations.
\end{proposition}

\begin{proof} The proof amounts to showing that
$Q^\alpha_i$ in \eqref{E:confsymm1} satisfy the 
determining equations \eqref{E:evolsymmeq} for 
a symmetry. This is a standard computation based 
on equation \eqref{E:productrulesimple} and on 
elementary properties of conformal Killing vectors 
and will therefore be omitted.
\end{proof}

Let $\fg=\fg_1+\cdots +\fg_n$ the decomposition of
a semi-simple Lie algebra into simple ideals.
Order the ideals so that for some
$0\leq p\leq n+1$, the ideals $\fg_m$, 
$m<p$, are real forms of simple complex 
Lie algebras and the ideals $\fg_m$, $m \geq p$, 
are realifications of simple complex Lie 
algebras with complex structures $J_m$. 
Let $P_m : \fg\to \fg_m$ be the projection induced 
by the above decomposition. Write $F^\alpha_{m,ij}$ 
for the Yang-Mills field tensor for $\fg_m$-valued fields.

Let $\xi^j$, $\tau^j$ be conformal Killing vectors 
and let $Q_m[\xi]$, $1\leq m\leq n$, ${Q}_{J,m}[\tau]$, 
$p\leq m\leq n$, be evolutionary vector fields 
on $J^1(A)$ with components $Q^\alpha_{m,i}[\xi]$, 
${Q}^\alpha_{J,m,i}[\tau]$ determined by 
\begin{equation}
\label{E:confJsymm}
P_lQ_{m,i}[\xi] = 
\delta_{lm}\xi^jF_{m,ij},\quad
P_l Q_{J,m,i}[\tau] = 
\delta_{lm}\tau^jJ_{m} F_{m,ij},\qquad 1\leq l\leq n.
\end{equation}
\begin{corollary} 
\label{C:confsymm}
Let $\xi^j$, $\tau^j$ be
conformal Killing vectors. Then the vector fields
$Q_m[\xi]$, $1\leq m\leq n$, ${Q}_{J,m}[\tau]$, 
$p\leq m\leq n$, are first order generalized
symmetries of the Yang-Mills equations
for $\fg$-valued fields.
\end{corollary}
\begin{proof} The proof of the Corollary is an
immediate consequence of Propositions 
\ref{P:adjointcommute} and \ref{P:confsymm1}.
\end{proof}

Due to the decomposition
$\fg=\fg_1+\cdots+\fg_n$ of $\fg$ into a direct
sum of simple ideals, the $\fg$-valued Yang-Mills 
equations decouple into system of $\fg_m$-valued
Yang-Mills equations, $1\leq m\leq n$.
Corollary \ref{C:confsymm} simply states 
that the conformal symmetries and their 
images under a complex structure of the 
component equations in this system yield 
symmetries of the full system. Theorem
\ref{T:maintheorem} assert that any generalized
symmetry of the $\fg$-valued Yang-Mills equations 
is equivalent to a sum of the symmetries
$Q_m[\xi]$, ${Q}_{J,m}[\tau]$. 

Spinorial methods play a crucial role in our 
analysis of symmetries of the Yang-Mills fields.
Given a tensorial object $T_{i_1\cdots i_p}$,
we write $T_{I_1I_1'\cdots I_pI_p'}$ for its 
spinor representative, where
$$
T_{I_1I_1'\cdots I_pI_p'}=
\sigma^{i_1}_{I_1I_1'}\cdots\sigma^{i_p}_{I_pI_{p'}}
T_{i_1\cdots i_p}.
$$
Here, apart from a constant factor, the matrices
$\sigma_{II'}^i$ are the identity matrix and the
Pauli spin matrices. 
We use bar to denote complex
conjugation and we raise and lower spinor indices
using the spin metric $\e_{IJ}=\e_{[IJ]}$, $\e_{01}=1$,
and its complex conjugate $\e_{I'J'}$. Fore more details,
see \cite{Penrose84}. Accordingly, we write
$$
\partial_{II'} = \sigma_{II'}^i\partial_i,\quad
D_{II'} = \sigma_{II'}^iD_i,\quad
\nabla_{II'} = \sigma_{II'}^i\nabla_i
$$
for the spinor representatives of the partial,
the total, and the covariant derivative operators.
Note that in spinor form equation \eqref{E:covarcommute} 
becomes
\begin{equation}
\label{E:covarcommutespin}
\nabla_{II'}\nabla_{JJ'}G-\nabla_{JJ'}\nabla_{II'}G=
\e_{I'J'}[\Phi_{IJ},G]+\e_{IJ}[\overline\Phi_{I'J'},G].
\end{equation}

Let
$\Phi^\alpha_{IJ;K_1K_1'\cdots K_{p}K_{p}'} = 
\Phi^\alpha_{(IJ);K_1K_1'\cdots K_{p}K_{p}'}$, 
$p\geq 0$,
be the spinorial variables determined by the equations

\begin{align}
\sigma^i_{II'}\sigma^j_{JJ'}\sigma^{k_1}_{K_1K_1'}
\cdots\sigma^{k_p}_{K_pK_p'}
&\nabla_{k_1}\cdots\nabla_{k_p}F^\alpha_{ij}=\nonumber\\
&\e_{I'J'}\Phi^\alpha_{IJ;K_1K_1'\cdots K_pK_p'} +
\e_{IJ}\overline\Phi^\alpha_{I'J';K_1K_1'\cdots K_pK_p'}.
\nonumber
\end{align}

Then, in spinor form, the Yang-Mills 
equations \eqref{E:YMequations} reduce to
\[
\nabla_{I'}^J\Phi_{IJ} = 0,
\]
while the determining equations for symmetries 
\eqref{E:evolsymmeq} become
\begin{equation}
\label{E:YMsymmspinor}
\nabla^{JJ'}\nabla_{II'}Q_{JJ'}-
\nabla^{JJ'}\nabla_{JJ'}Q_{II'} -
[\Phi_I^J,Q_{JI'}]-[\overline\Phi_{I'}^{J'},Q_{IJ'}]=0
\quad\mbox{on $\mathcal R^p$}.
\end{equation}

Next define symmetrized variables 
$a^\alpha_{i_1i_2\cdots i_{p+1}}$,
$\Phi^\alpha_{IJ}{}_{K_1\cdots K_p}^
{K_1'\cdots K_p'}$, $p\geq0$, by
\begin{equation}
\label{E:symmvariables}
a^\alpha_{i_1i_2\cdots i_{p+1}} = 
a^\alpha_{(i_1,i_2\cdots i_{p+1})},\quad
\Phi^\alpha{}_{IJK_1\cdots K_p}^{\hp{IJ}K_1'\cdots K_p'}=
\Phi^\alpha{}_{(IJ;K_1\cdots K_{p})}^{\hp{IJ;}(K_1'\cdots K_p')},
\end{equation}
where round brackets indicate symmetrization 
in the enclosed indices.

In order to avoid excessive proliferation of 
indices we will streamline our notation by 
employing multi-indices of integers to designate 
groups of indices in which an object is symmetric. 
In the case of the space-time indices, we denote 
multi-indices by boldface lower case letters, and 
in the case of spinorial indices, by boldface capital 
letters. Hence, for example, 
$\mathbf i_p = (i_1,i_2,\dots, i_p)$, where each $i_j$ 
is either $0$, $1$, $2$, or $3$, and 
$\mathbf K'_p = (K_1',K_2',\dots,K_p')$, where each 
$K_j'$ is either $0$ or $1$. We also combine
multi-indices by the rule 
$\mathbf i_pi_{p+1} = \mathbf i_{p+1}$. Accordingly, 
we will write
\[
a^\alpha_{\mathbf i_p} = a^\alpha_{i_1\cdots i_p},\qquad
\Phi^\alpha{}_{\mathbf K_{p+2}}^{\mathbf K_p'}=
\Phi^\alpha_{K_{p+2}K_{p+1}}{}_{K_1\cdots K_p}^
{K_1'\cdots K_p'}.
\]
We will, moreover, collectively designate the variables
$a^\alpha_{\mathbf i_{p+1}}$, 
$\Phi^\alpha{}^{\mathbf K_p'}_{\mathbf K_{p+2}}$ 
by $\partial^pa$, $\partial^p\Phi$, and we let
$a^{[p]}$, $\Phi^{[p]}$ stand for the variables
$\partial^q a$, $\partial^q\Phi$, $0\leq q\leq p$.
Note that the variables $\partial^p\Phi$
are of order $p+1$.
 
The proof of the following result appears in \cite{Torre95}.
\begin{proposition}
\label{P:solcoordinates}
The variables
$$
x^j,\quad a^{[p+2]},\quad\Phi^{[p+1]},
$$
form a coordinate system on the $p$-fold prolonged 
solution manifold $\mathcal R^p$, $p\geq0$.
\end{proposition}
Thus, in particular, any symmetry of the Yang-Mills
equations of order $p\geq1$ is equivalent to one depending 
on the variables $x^j$, $a^{[p]}$, $\Phi^{[p-1]}$ only. 

The following result is pivotal in our analysis of 
symmetries of the Yang-Mills equations. The proof
of the first part of the Proposition is based on standard 
index manipulations and will be omitted. The second 
part, however, relies on a lengthy computation 
and we will therefore defer its proof to the Appendix.

\begin{proposition}
\label{P:dersymmvar} {\rm (i)} The derivative 
$D_{i_{p+1}}\ader{p}$ of the symmetrized
variable $\ader{p}$ satisfies the equation
\begin{equation}
\label{E:derasymm}
D_{i_{p+1}}a^\alpha_{i_1\cdots i_{p}} = 
a^\alpha_{i_1\cdots i_{p}i_{p+1}}-
\frac{1}{p+1}\nabla_{(i_1}\cdots\nabla_{i_{p-1}}
F^\alpha_{i_{p})i_{p+1}}+b^\alpha_{i_1\cdots i_{p}i_{p+1}},
\end{equation}
where the functions $b^\alpha_{i_1\cdots i_{p}i_{p+1}}$
are of order $p-1$.\newline
{\rm (ii)} When restricted to the solution manifold 
$\mathcal R^{p}$, the covariant derivative
$\nabla_{K_{p+3}}^{K_{p+1}'}\Pvar{K}{\alpha}{p+2}{p}$ 
of the symmetrized variable
$\Pvar{K}{\alpha}{p+2}{p}$, $p\geq1$, 
satisfies the equation
\begin{eqnarray}
\nabla_{K_{p+3}}^{K_{p+1}'}\Pvar{K}{\alpha}{p+2}{p} &&= 
\Pvar{K}{\alpha}{p+3}{p+1}+
\frac{p^2+p-2}{2(p+1)}\e^{K_{p+1}'(K_p'}[\Phi_{(K_{p+3}K_{p+2}},
\Pvar{K}{}{p+1}{p-1}{}^)_)]^\alpha-\nonumber\\
&&\frac{3(1-\delta_{1p})}{p+3}\e_{K_{p+3}(K_{p+2}}[\Phi_{K_{p+1}K_{p}},
\overline\Phi^{\mathbf K_{p+1}'}_{\mathbf K_{p-1})}]^\alpha+\nonumber\\
&&\frac{p(p+5)}{2(p+3)}\e_{K_{p+3}(K_{p+2}}
[\Phibar^{(K_{p+1}'K_p'},
\Phi_{\mathbf K_{p+1})}^{\mathbf K_{p-1}')}]^\alpha+
\label{E:derPhisymmvar}\\
&&(\frac{(p+1)(p+2)}{p+3}-\frac{3}{4}\delta_{1p})
\e_{K_{p+3}(K_{p+2}}\e^{K_{p+1}'(K_p'}
[\Phi_{|S|K_{p+1}},
\Phi_{\mathbf K_{p})}^{\mathbf K_{p-1}')S}]^\alpha+
\nonumber\\
&&\frac{p(p-1)}{p+1}\e^{K_{p+1}'(K_p'}\e_{K_{p+3}(K_{p+2}}
[\Phibar_{|S'|}{}^{K_{p-1}'},
\Phi_{\mathbf K_{p+1})}^{\mathbf K_{p-2}')S'}]^\alpha
+\Psi_{1,}^\alpha{}_{\mathbf K_{p+3}}^{\mathbf K_{p+1}'},\nonumber
\end{eqnarray}
where $\Psi_{1,}^\alpha{}_{\mathbf K_{p+3}}^{\mathbf K_{p+1}'}=
\Psi_{1,}^\alpha{}_{\mathbf K_{p+3}}^{\mathbf K_{p+1}'}(\Phi^{[p-2]})$
only depend on the variables $\Phi^{[p-2]}$.  
Moreover, on $\mathcal R^{[p+1]}$, we have that
\begin{equation}
\label{E:wavePhiOrdp}
\nabla^2 \Phi^\alpha{}^{\mathbf K_p'}_{\mathbf K_{p+2}}=
2(p+2-\delta_{0p})[\Phi_{S(K_{p+2}},
\Phi_{\mathbf K_{p+1})}^{\mathbf K_p'S}]^\alpha+
2p[\overline\Phi_{S'}^{(K_p'},
\Phi^{\mathbf K_{p-1}')S'}_{\mathbf K_{p+2}}]^\alpha+
\Psi_{2,}^\alpha{}^{\mathbf K_p'}_{\mathbf K_{p+2}},
\end{equation}
where $\Psi_{2,}^\alpha{}^{\mathbf K_p'}_{\mathbf K_{p+2}}=
\Psi_{2,}^\alpha{}^{\mathbf K_p'}_{\mathbf K_{p+2}}(\Phi^{[p-1]})$
only depend on the variables $\Phi^{[p-1]}$.

\end{proposition}

Next write
\[
u^\alpha{}_{\mathbf K_{p+2}}^{\mathbf K_p'} = 
\mbox{Re}\,\Phi^\alpha{}_{\mathbf K_{p+2}}^{\mathbf K_p'},\quad
v^\alpha{}_{\mathbf K_{p+2}}^{\mathbf K_p'}=
\mbox{Im}\,\Phi^\alpha{}_{\mathbf K_{p+2}}^{\mathbf K_p'}
\]
for the real and imaginary parts of the symmetrized
variables $\Phi^\alpha{}_{\mathbf K_{p+2}}^{\mathbf K_p'}$
Suppose that a differential function $G$ only depends
on the variables $x^j$, $a^{[p]}$
$u^{[p-1]}$, $v^{[p-1]}$.
We can always assume that this is the case with a 
symmetry of the Yang-Mills equations of order $p$. 
We write
\[
\partial_{a^\alpha}^{\mathbf i_{q}}G,\quad
\partial_{u^\alpha}{}^{\mathbf K_{q+2}}_{\mathbf K_q'}G,\quad
\partial_{v^\alpha}{}^{\mathbf K_{q+2}}_{\mathbf K_q'}G,
\]
for the weighted partial derivatives of $G$ with 
respect to the variables $a^\alpha_{\mathbf i_q}$,
$u^\alpha{}_{\mathbf K_{q+2}}^{\mathbf K_q'}$,
$v^\alpha{}_{\mathbf K_{q+2}}^{\mathbf K_q'}$.
Define
\[
\Pder{\alpha}{\mathbf K_{q+2}}{\mathbf K_q'} = \frac{1}{2}
(\partial_{u^\alpha}{}^{\mathbf K_{q+2}}_{\mathbf K_q'}-
\mbox{i}\partial_{v^\alpha}{}^{\mathbf K_{q+2}}_{\mathbf K_q'}),\qquad
\Pderbar{\alpha}{\mathbf K_{q+2}'}{\mathbf K_q}=
\overline{\Pder{\alpha}{\mathbf K_{q+2}}{\mathbf K_q'}},
\]
where $\mbox{i}$ is the imaginary unit.
Thus, in particular,
\jot10pt
\begin{eqnarray}
\partial_{a^\alpha}^{\mathbf i_q}a^\beta_{\mathbf j_r}&&=
\begin{cases} 
\delta^\beta_\alpha{}\delta_{j_1}^{(i_1}\cdots\delta_{j_q}^{i_q)},
&\mbox{if $q=r$},\\
 0,&\mbox{if $q\neq r$};\end{cases}\nonumber\\
\Pder{\alpha}{\mathbf K_{q+2}}{\mathbf K_q'}
\Pvar{L}{\beta}{r+2}{r}&&=
\begin{cases}
\delta^\beta_\alpha
\e_{L_1}{}^{(K_1}\cdots\e_{L_{q+2}}{}^{K_{q+2})}
\e_{(K_1'}{}^{L_1'}\cdots\e_{K_q')}{}^{L_q'},
&\mbox{if $q=r$,}\\
0,&\mbox{if $q\neq r$};\end{cases}\label{E:pardersymmvar}\\
\Pder{\alpha}{\mathbf K_{q+2}}{\mathbf K_q'}
\Pvarbar{L}{\beta}{r+2}{r}&&=\;\;\;0.\nonumber
\end{eqnarray}

\section{Generalized symmetries of the Yang-Mills equations}
\label{S:proof}
In this Section we prove Theorem 
\ref{T:maintheorem} by completely 
classifying generalized symmetries 
of the Yang-Mills equations. Let 
\[
\wQ^\alpha_i = \wQ^\alpha_i(x^j, a^{[p]}, \Phi^{[p-1]}),
\]
be the components of a generalized symmetry $\wQ$ 
of order $p$ of the equations, which, without 
loss of generality, we assume to be a function 
of $x^j$ and the symmetrized variables 
$a^{[p]}$, $\Phi^{[p-1]}$ only.  

We start by analyzing the highest order terms in the 
symmetrized variables $\partial^q a$ in the 
determining equations for $\wQ$ to show 
that $\wQ$ is equivalent to a 
symmetry that does not depend on 
$\partial^qa$, $q\geq 1$. The proof of the 
following Proposition is a straightforward
computation and will be omitted.

\begin{proposition}
\label{P:reduction}
Let 
$T_{\alpha}{}_{i}^{k_1\cdots k_p} = 
T_{\alpha}{}_{i}^{(k_1\cdots k_p)}$
be some constants satisfying the equations
$$
T_{\alpha}{}_{i}^{k_1\cdots k_p}
a^\alpha_{k_1\cdots k_p}{}_j^j = 
T_{\alpha}{}_{j}^{k_1\cdots k_p}
a^\alpha_{k_1\cdots k_p}{}_i^j.
$$
Then there are constants 
$S_\alpha^{k_1\cdots k_{p-1}}=
S_\alpha^{(k_1\cdots k_{p-1})}$
so that 
\[
T_{\alpha}{}_{i}^{k_1\cdots k_p} = 
\delta_i^{(k_1}S_\alpha^{k_2\cdots k_p)}.
\]
\end{proposition}

\begin{lemma}
\label{L:aReduction}
Let 
\[
\wQ^\alpha_i = 
\wQ^\alpha_i(x^j,a^{[p]},\Phi^{[p-1]}),\qquad p\geq0,
\]
be the components of a symmetry $\wQ$ of the Yang-Mills 
equations of order $p$. Then $\wQ$ is equivalent
to a symmetry $Q$ with components $Q^\alpha_i$
of the form
\begin{equation}
\label{E:areduced}
Q^\alpha_i =r^\alpha_i(x^j,\Phi^{[2p-1]})+
q^\alpha_\beta(x^j,\Phi^{[2p-1]})a^\beta_i.
\end{equation}
\end{lemma} 

\begin{proof}
Substitute $\wQ^\alpha_i$ in the 
determining equations \eqref{E:evolsymmeq} 
and collect the coefficients of the 
terms $\partial^{p+2}a$. On account 
of Proposition \ref{P:dersymmvar}, 
this yields the equation
\[
(\partial_{a^\beta}^{\mathbf k_{p+1}} \wQ^\alpha_{j})
a^\beta_{\mathbf k_{p+1}i}{}^j-
(\partial_{a^\beta}^{\mathbf k_{p+1}}\wQ^{\alpha}_i)
 a^\beta_{\mathbf k_{p+1}j}{}^j=0.
\]
Thus, by Proposition \ref{P:reduction}, 
there are smooth differential functions
$q_\beta^{\alpha,\mathbf k_p} =
q_\beta^{\alpha,\mathbf k_{p}}(x^j,a^{[p]},\Phi^{[p-1]})$
such that
\begin{equation}
\label{E:ordpa}
\partial_{a^\beta}^{\mathbf k_{p+1}}\wQ_i^{\alpha} = 
\delta_i^{(k_{p+1}}q_\beta^{|\alpha|,\mathbf k_{p})}.
\end{equation}

After differentiation, equation \eqref{E:ordpa} yields
\begin{equation}
\label{E:ordpatwice}
\delta_i^{(k_{p+1}}\partial_{a^\gamma}^{|\mathbf l_{p+1}}
q_\beta^{\alpha|,\mathbf k_{p})}=
\delta_i^{(l_{p+1}}\partial_{a^\beta}^{|\mathbf k_{p+1}}
q_\gamma^{\alpha|,\mathbf l_{p})}.
\end{equation}
Multiply \eqref{E:ordpatwice} by 
$X_{\mathbf k_{p+1}}$, $Y_{\mathbf l_{p+1}}$
to see that
\[
(\partial_{a^\gamma}^{\mathbf l_{p+1}}
q_\beta^{\alpha,\mathbf k_{p}}) 
X_{\mathbf k_{p}}Y_{\mathbf l_{p+1}}X_i=
(\partial_{a^\beta}^{\mathbf k_{p+1}}
q_\gamma^{\alpha,\mathbf l_{p}})
X_{\mathbf k_{p+1}}Y_{\mathbf l_{p}}Y_i,
\]
which implies that 
$\partial_{a^\gamma}^{\mathbf l_{p+1}}
q_\beta^{\alpha,\mathbf k_{p}} 
X_{\mathbf k_{p}}Y_{\mathbf l_{p+1}}=0$
whenever $X_i$, $Y_i$ are linearly independent, 
and thus, by continuity,  
\[\partial_{a^\gamma}^{\mathbf l_{p+1}}
q_\beta^{\alpha,\mathbf k_{p}}=0.\]
Hence
$q_\beta^{\alpha,\mathbf k_p} = 
q_\beta^{\alpha,\mathbf k_{p}}(x^j,a^{[p-1]},\Phi^{[p-1]})$, 
and consequently, by \eqref{E:ordpa}, we have that
\begin{equation}
\label{E:ordpaterms}
\wQ^\alpha_i = q_\beta^{\alpha,\mathbf k_{p}}(x^j,a^{[p-1]},
\Phi^{[p-1]})a^\beta_{\mathbf k_{p}i}+
t^\alpha_i(x^j,a^{[p-1]},\Phi^{[p-1]}),
\end{equation}
for some functions 
$t^\alpha_i=t^\alpha_i(x^j,a^{[p-1]},\Phi^{[p-1]})$.

If $p=0$, the above arguments show 
that \eqref{E:areduced} holds. 
Suppose that $p\geq1$. Next we collect 
terms in the determining equations 
for $\wQ$ involving products of
$\partial^{p}a$, $\partial^{p+1}a$. Since $p\geq1$,
we see from Proposition \ref{P:dersymmvar} that these
only arise from the derivatives of $\wQ^\alpha_i$
with respect to the variables $\partial^{p-1}a$,
$\partial^pa$. 

After some manipulations we obtain the equation
\begin{equation}
(\partial_{a^\gamma}^{\mathbf l_p}q_\beta^{\alpha,\mathbf k_{p}}-
\partial_{a^\beta}^{\mathbf k_p}q_\gamma^{\alpha,\mathbf l_{p}})
(a^\beta_{\mathbf k_pj}a^\gamma_{\mathbf l_p}{}^j_i+
a^\gamma_{\mathbf l_pi}a^\beta_{\mathbf k_p}{}^j_j)=0.
\end{equation}
Now apply the operators 
$A^p_{\rv,X}$ and $A^{p+1}_{\rw,Y}$, where 
\[
A^q_{\rv,X} = \rv^\alpha X_{\mathbf i_q}
\partial_{a^\alpha}^{\mathbf i_q},
\] 
to the above equation to see that
\[
(\partial_{a^\gamma}^{\mathbf l_p}q_\beta^{\alpha,\mathbf k_{p}}-
\partial_{a^\beta}^{\mathbf k_p}q_\gamma^{\alpha,\mathbf l_{p}})
(\rv^\beta \rw^\gamma X_{\mathbf k_pj}Y_{\mathbf l_pi}{}^j+
\rv^\gamma \rw^\beta X_{\mathbf l_pi} Y_{\mathbf k_pj}{}^j)=0.
\]
This implies that
\begin{equation}
\label{E:derofqred}
(\partial_{a^\gamma}^{\mathbf l_p}q_\beta^{\alpha,\mathbf k_{p}}-
\partial_{a^\beta}^{\mathbf k_p}q_\gamma^{\alpha,\mathbf l_{p}})
X_{\mathbf k_p}Y_{\mathbf l_p}=0,
\end{equation}
whenever $X_i$, $Y_i$ are linearly 
independent and $X_jY^j\neq0$. 
By continuity, equation \eqref{E:derofqred}
holds for all 
$X_i$, $Y_i$, and hence, the
functions $q_\beta^{\alpha,\mathbf k_p}$
satisfy the integrability conditions
\begin{equation}
\label{E:qintcond}
\partial_{a^\gamma}^{\mathbf l_p}
q_\beta^{\alpha,\mathbf k_{p}}=
\partial_{a^\beta}^{\mathbf k_p}
q_\gamma^{\alpha,\mathbf l_{p}}.
\end{equation}

Consequently, if we define a $\fg$-valued 
differential function $\mathcal X^\alpha$ by
\[
\mathcal X^\alpha(x^j,a^{[p-1]},\Phi^{[p-1]}) = 
\int_0^1 q_\beta^{\alpha,\mathbf k_{p}}(x^j,a^{[p-2]},
t\partial^{p-1}a,\Phi^{[p-1]}) a^\beta_{\mathbf k_{p}}dt,
\]
then, by \eqref{E:ordpaterms}, 
\eqref{E:qintcond}, the difference 
$\wQ^\alpha_i - \nabla_i\mathcal X^\alpha$,
when restricted to $\mathcal R^{[p-1]}$, 
only involves the variables 
$\partial^q a$ up to order $p-1$. 
Thus $\wQ$ is equivalent to 
a symmetry $\widehat Q$ with components 
\[
\widehat Q^\alpha_i = 
\widehat Q^\alpha_i(x^j,a^{[p-1]},\Phi^{[p]}).
\]
Now one can inductively repeat the above 
argument to conclude that $\wQ$ 
is equivalent to a symmetry $Q$
with components $Q^\alpha_{i}$
as in \eqref{E:areduced}.
\end{proof}

Next suppose that we have a symmetry 
$\wQ$ with components 
$\wQ^\alpha_{II'}$ given by
\begin{equation}
\label{E:areducedp}
\wQ_{II'}^\alpha =  
\tr^\alpha_{II'}(x^j,\Phi^{[p]})+
\tilde q^\alpha_\beta(x^j,\Phi^{[p]})a^\beta_{II'},
\qquad p\geq 1.
\end{equation}
In the following Lemma we analyze terms
in the determining equations for $\wQ$
involving the variables 
$\partial^{p+1}\Phi$, $\partial^{p+2}\Phi$.
We use the symbol c.c. to denote the complex 
conjugates of the terms preceding it in an 
expression.

\begin{lemma}\label{L:ordpPhired}
Let $\wQ$ be a symmetry of the Yang-Mills
equations with components $\wQ^\alpha_{II'}$ 
as in \eqref{E:areducedp}. Then $\wQ$ is 
equivalent to a symmetry $Q$ with components
$Q_{II'}^\alpha$ given by
\begin{eqnarray}
&&Q_{II'}^\alpha =
s^\alpha_\beta{}_{I'\mathbf K_p'}^{\mathbf K_{p+1}}(x^j)
\Phi^\beta{}_{I\mathbf K_{p+1}}^{\mathbf K_{p}'}+
t^\alpha_\beta{}_{I\mathbf K_{p-1}'}^{\mathbf K_{p+2}}(x^j)
\Phi^\beta{}_{I'\mathbf K_{p+2}}^{\mathbf K_{p-1}'}+\nonumber\\
&&\quad w^\alpha_\beta{}^{\mathbf K_{p+1}}_{II'\mathbf K_{p-1}'}(x^j)
\Pvar{K}{\beta}{p+1}{p-1}+ c.c.+
q^\alpha_\beta(x^j,\Phi^{[p-2]})a^\beta_{II'}+
v^\alpha_{II'}(x^j,\Phi^{[p-1]}),\label{E:ordpQ}
\end{eqnarray}
where 
$s^\alpha_\beta{}_{\mathbf K_{p+1}'}^{\mathbf K_{p+1}}$,
$t^\alpha_\beta{}_{\mathbf K_{p-1}'}^{\mathbf K_{p+3}}$
are Killing spinors and satisfy the equations
\begin{equation} 
\label{E:stadjcomm}
c^\alpha_{\beta\gamma}
s^\beta_\delta{}^{\mathbf K_{p+1}}_{\mathbf K_{p+1}'}+
s^\alpha_\beta{}^{\mathbf K_{p+1}}_{\mathbf K_{p+1}'}
c^\beta_{\gamma\delta}=0,
\qquad
c^\alpha_{\beta\gamma}
t^\beta_\delta{}^{\mathbf K_{p+3}}_{\mathbf K_{p-1}'}+
t^\alpha_\beta{}^{\mathbf K_{p+3}}_{\mathbf K_{p-1}'}
c^\beta_{\gamma\delta} =0,
\end{equation}
$w^\alpha_\beta{}^{\mathbf K_{p+2}}_{\mathbf K_{p}'}$ 
are given by
\begin{equation}
\label{E:wdefn}
w^\alpha_\beta{}^{\mathbf K_{p+2}}_{\mathbf K_{p}'}=
-\frac{p}{p+2}\partial^{(K_{p+2}}_{P'}
s^{|\alpha|}_\beta{}_{\mathbf K_p'}^{\mathbf K_{p+1})P'}+
\frac{p+2}{p+4}\partial_{P(K_p'}
t^\alpha_{|\beta|}{}^{\mathbf K_{p+2}P}_{\mathbf K_{p-1}')},
\end{equation}
and where
\begin{equation}
\label{E:vsymmder}
\Pder{\gamma}{(\mathbf K_{p+1}}{(\mathbf K_{p-1}'}
v^{|\alpha|}{}^{K_{p+2})}_{K_{p}')} = 0,\qquad
\Pderbar{\gamma}{(\mathbf K_{p+1}'}{(\mathbf K_{p-1}}
v^{|\alpha|}{}^{K_{p+2}')}_{K_{p})} = 0.
\end{equation}
\end{lemma}
\begin{proof}
We substitute $\wQ^\alpha_{II'}$ in 
\eqref{E:areducedp} into the determining 
equations \eqref{E:YMsymmspinor} and use 
Proposition \ref{P:dersymmvar} to conclude 
that in the resulting equations terms 
involving the variables $\partial^{p+2}\Phi$ 
yield the expression
\begin{align}
(\Pder{\gamma}{\mathbf K_{p+2}}{\mathbf K_p'}\tr^\alpha_{JJ'})
&\Phi^\gamma{}_{II'\mathbf K_{p+2}}^{JJ'\mathbf K_p'}
+
(\Pderbar{\gamma}{\mathbf K_{p+2}'}{\mathbf K_p}\tr^\alpha_{JJ'})
\overline{\Phi}^\gamma{}_{II'\mathbf K_{p+2}'}^{JJ'\mathbf K_p}+\nonumber\\
&\big((\Pder{\gamma}{\mathbf K_{p+2}}{\mathbf K_p'}
\tq^\alpha_\beta)\,
\Phi^\gamma{}_{II'\mathbf K_{p+2}}^{JJ'\mathbf K_p'}+
(\Pderbar{\gamma}{\mathbf K_{p+2}'}{\mathbf K_p}
\tq^\alpha_\beta)\,
\overline{\Phi}^\gamma{}_{II'\mathbf K_{p+2}'}^{JJ'\mathbf K_p}
\big)
a^\beta_{JJ'},\nonumber
\end{align}
which must vanish. Hence we have that
\begin{equation}
\label{E:ordp2Phi}
\tq^\alpha_\beta = \tq^\alpha_\beta(x^j,\Phi^{[p-1]}),\quad
\Pder{\gamma}{(\mathbf K_{p+2}}{(\mathbf K_p'}
\tr^{|\alpha|}{}^{K_{p+3})}_{K_{p+1}')}=0,\quad
\Pderbar{\gamma}{(\mathbf K_{p+2}'}{(\mathbf K_p}
\tr^{|\alpha|}{}^{K_{p+3}')}_{K_{p+1})}=0.
\end{equation}

Next we collect terms quadratic in the 
variables $\partial^{p+1}\Phi$,
$\partial^{p+1}\overline\Phi$  in the determining
equations for $\wQ$. By \eqref{E:ordp2Phi} 
these only arise from the term 
$\tr^\alpha_{II'}$ in $\wQ^\alpha_{II'}$.
Note that the derivative 
$\nabla^{JJ'}\nabla_{II'}\wQ^\alpha_{JJ'}$,
when restricted to $\mathcal R^p$, yields the 
following quadratic terms of order $p+2$,
\begin{align}
\bigl((\Pder{\beta}{\mathbf K_{p+2}}{\mathbf K_{p}'}
\Pder{\gamma}{\mathbf L_{p+2}}{\mathbf L_{p}'}
&\tr^\alpha_{JJ'})
\Phi^\gamma{}^{\mathbf L_p'JJ'}_{\mathbf L_{p+2}}+
(\Pder{\beta}{\mathbf K_{p+2}}{\mathbf K_{p}'}
\Pderbar{\gamma}{\mathbf L_{p+2}'}{\mathbf L_{p}}
\tr^\alpha_{JJ'})
\overline{\Phi}^\gamma{}^{\mathbf L_pJJ'}_{\mathbf L_{p+2}'})
\Pvar{K}{\beta}{{p+2}}{p}{}_{II'}+\nonumber\\
\bigl((\Pderbar{\beta}{\mathbf K_{p+2}'}{\mathbf K_{p}}
\Pder{\gamma}{\mathbf L_{p+2}}{\mathbf L_{p}'}
&\tr^\alpha_{JJ'})
\Phi^\gamma{}^{\mathbf L_p'JJ'}_{\mathbf L_{p+2}}+
(\Pderbar{\beta}{\mathbf K_{p+2}'}{\mathbf K_{p}}
\Pderbar{\gamma}{\mathbf L_{p+2}'}{\mathbf L_{p}}
\tr^\alpha_{JJ'})
\overline{\Phi}^\gamma{}^{\mathbf L_pJJ'}_{\mathbf L_{p+2}'}\bigr)
\Pvarbar{K}{\beta}{{p+2}}{p}{}_{II'},\nonumber
\end{align}
all of which vanish due to \eqref{E:ordp2Phi}.
Thus the terms quadratic in $\partial^{p+1}\Phi$,
$\partial^{p+1}\overline\Phi$ in the determining
equations for $\wQ$ yield the equation
\begin{align}
(\Pder{\beta}{\mathbf K_{p+2}}{\mathbf K_{p}'}
&\Pder{\gamma}{\mathbf L_{p+2}}{\mathbf L_{p}'}
\tr^\alpha_{II'})
\Pvar{K}{\beta}{{p+2}}{p}{}_{JJ'}
\Phi^\gamma{}_{\mathbf L_{p+2}}^{\mathbf L_p'JJ'}+\nonumber\\
&2(\Pder{\beta}{\mathbf K_{p+2}}{\mathbf K_{p}'}
\Pderbar{\gamma}{\mathbf L_{p+2}'}{\mathbf L_{p}}
\tr^\alpha_{II'})
\Pvar{K}{\beta}{{p+2}}{p}{}_{JJ'}
\overline{\Phi}^\gamma{}_{\mathbf L_{p+2}'}^{\mathbf L_pJJ'}+
\nonumber\\
&\qquad\qquad(\Pderbar{\beta}{\mathbf K_{p+2}'}{\mathbf K_{p}}
\Pderbar{\gamma}{\mathbf L_{p+2}'}{\mathbf L_{p}}
\tr^\alpha_{II'})
\Pvarbar{K}{\beta}{{p+2}}{p}{}_{JJ'}
\overline{\Phi}^\gamma{}_{\mathbf L_{p+2}'}^{\mathbf L_pJJ'}=0,
\nonumber
\end{align}
from which it follows that
\begin{equation}
\label{E:scndderpr}
\Pder{\beta}{\mathbf K_{p+2}}{\mathbf K_p'}
\Pder{\gamma}{\mathbf L_{p+2}}{\mathbf L_p'}
\tr^\alpha_{II'} =0,\quad
\Pder{\beta}{\mathbf K_{p+2}}{\mathbf K_p'}
\Pderbar{\gamma}{\mathbf L_{p+2}'}{\mathbf L_p}
\tr^\alpha_{II'} =0,\quad
\Pderbar{\beta}{\mathbf K_{p+2}'}{\mathbf K_p}
\Pderbar{\gamma}{\mathbf L_{p+2}'}{\mathbf L_p}
\tr^\alpha_{II'}=0,
\end{equation}
that is, the functions $\tr^\alpha_{II'}$ 
are linear in the highest order field 
variables. Hence by \eqref{E:ordp2Phi},
\eqref{E:scndderpr},
the components $\wQ^\alpha_{II'}$ reduce to
\begin{align}
\wQ_{II'}^\alpha = 
s^\alpha_\beta{}_{I'\mathbf K_p'}^{\mathbf K_{p+1}}&(x^j,\Phi^{[p-1]})
\Phi^\beta{}_{I\mathbf K_{p+1}}^{\mathbf K_{p}'}+
t^\alpha_\beta{}_{I\mathbf K_{p-1}'}^{\mathbf K_{p+2}}(x^j,\Phi^{[p-1]})
\Phi^\beta{}_{I'\mathbf K_{p+2}}^{\mathbf K_{p-1}'}+\nonumber\\
&u^\alpha_\beta{}_{\mathbf K_{p-1}'}^{\mathbf K_{p+1}}(x^j,\Phi^{[p-1]})
\Phi^\beta{}_{II'\mathbf K_{p+1}}^{\mathbf K_{p-1}'}+ c.c.+\label{E:ordpQ2}\\
&\qquad\qquad\qquad\tq^\alpha_\beta(x^j,\Phi^{[p-1]}) a^\beta_{II'} +
\tilde v^\alpha_{II'}(x^j,\Phi^{[p-1]}),\nonumber
\end{align}
where
$s^\alpha_\beta{}_{\mathbf K_{p+1}'}^{\mathbf K_{p+1}}$,
$t^\alpha_\beta{}_{\mathbf K_{p-1}'}^{\mathbf K_{p+3}}$,
$u^\alpha_\beta{}_{\mathbf K_{p-1}'}^{\mathbf K_{p+1}}$
are symmetric in their spinorial indices.

We next analyze terms linear in the variables 
$\partial^{p+1}\Phi$ in the determining equations
for $\wQ$. Write \eqref{E:ordpQ2} as 
\[
\wQ_{II'}^\alpha = 
r^\alpha_\beta{}^{\hp{II'}\mathbf K_{p+2}}_{II'\mathbf K_p'}
\Pvar{K}{\beta}{p+2}{p} +
\overline{r}^\alpha_\beta{}^{\hp{II'}\mathbf K_{p+2}'}_{II'\mathbf K_p}
\Pvarbar{K}{\beta}{p+2}{p} +\tq^\alpha_\beta\, a^\beta_{II'}+
\tilde v^\alpha_{II'},
\]
so that 
\begin{equation}
\label{E:rsymmetrization}
r^\alpha_\beta{}^{(I}_{(I'}{}^{\mathbf K_{p+2})}_{\mathbf K_p')}=0.
\end{equation}
For example, we use \eqref{E:productrule} to compute
\begin{eqnarray}
&&\nabla^{JJ'}\nabla_{II'}(
r^\alpha_\beta{}
^{\hp{JJ'}\mathbf K_{p+2}}_{JJ'\mathbf K_p'} \Pvar{K}{\beta}{p+2}{p})=
r^\alpha_\beta{}^{\hp{JJ'}\mathbf K_{p+2}}_{JJ'\mathbf K_p'}
\nabla^{JJ'}\nabla_{II'}\Pvar{K}{\beta}{p+2}{p}+\nonumber\\
&&(D^{JJ'}r^\alpha_\beta{}^{\hp{JJ'}\mathbf K_{p+2}}_{JJ'\mathbf K_p'})
\nabla_{II'}\Pvar{K}{\beta}{p+2}{p}+
(D_{II'}r^\alpha_\beta{}^{\hp{JJ'}\mathbf K_{p+2}}_{JJ'\mathbf K_p'})
\nabla^{JJ'}\Pvar{K}{\beta}{p+2}{p}+\nonumber\\
&&(D^{JJ'}D_{II'}r^\alpha_\beta{}^{\hp{JJ'}\mathbf K_{p+2}}_{JJ'\mathbf K_p'})
\Pvar{K}{\beta}{p+2}{p}-\label{E:excomputation1}\\
&&(c^\alpha_{\beta\gamma} 
r^\beta_\delta{}^{\hp{JJ'}\mathbf K_{p+2}}_{JJ'\mathbf K_p'}+
r^\alpha_\beta{}^{\hp{JJ'}\mathbf K_{p+2}}_{JJ'\mathbf K_p'}
c^\beta_{\gamma\delta})(a^\gamma{}^{JJ'}\nabla_{II'}\Pvar{K}{\delta}{p+2}{p}+
a^\gamma_{II'}\nabla^{JJ'}\Pvar{K}{\delta}{p+2}{p})+
\Upsilon^\alpha_{1,II'},\nonumber\\
&&\nabla^{JJ'}\nabla_{II'}(\tilde q^\alpha_\beta a^\beta_{JJ'})=
(D^{JJ'}D_{II'}\tilde q^\alpha_\beta) a^\beta_{JJ'}+
\Upsilon^\alpha_{2,II'},\label{E:excomputation2}
\end{eqnarray}
where $\Upsilon^\alpha_{1,II'}$,
$\Upsilon^\alpha_{2,II'}$ are of order $p+1$.

Recall that $p\geq1$. Apply the operator 
$B^{p+1}_{\rwseven,X,Y}$, where
\begin{equation}
\label{E:Boperator}
B^{q}_{\rwseven,X,Y} = \rw^\alpha X_{\mathbf K_{q+2}} Y^{\mathbf K_{q}'}
\partial_{\Phi^\alpha}{}^{\mathbf K_{q+2}}_{\mathbf K_{q}'},
\qquad q\geq0,
\end{equation}
to the determining equations for $\wQ$. 
With the help of equations \eqref{E:productrule}, 
\eqref{E:derPhisymmvar}, \eqref{E:pardersymmvar},
\eqref{E:excomputation1}, \eqref{E:excomputation2} 
we conclude that in the resulting equations
the terms not involving the variables 
$a^\alpha_{II'}$ yield the equation
\begin{eqnarray}
&&\big((\Pder{\gamma}{\mathbf K_{p+1}}{\mathbf K_{p-1}'}
r^\alpha_\beta{}^{\hp{JJ'}\mathbf L_{p+2}}_{JJ'\mathbf L_p'})
\Pvar{L}{\beta}{p+2}{p}+
\big(\Pder{\gamma}{\mathbf K_{p+1}}{\mathbf K_{p-1}'}
\overline{r}^\alpha_\beta{}^{\hp{JJ'}\mathbf L_{p+2}'}_{JJ'\mathbf L_p})
\Pvarbar{L}{\beta}{p+2}{p}\big)
X_{\mathbf K_{p+1}I}^{J} Y^{\mathbf K_{p-1}'J'}_{I'}+\nonumber\\
&&\qquad\quad(D^{JJ'}r^\alpha_\gamma{}^{\hp{JJ'}\mathbf K_{p+2}}_{JJ'\mathbf K_p'})
X_{\mathbf K_{p+2}I} Y^{\mathbf K_{p}'}_{I'}-
2(D^{JJ'}r^\alpha_\gamma{}^{\hp{II'}\mathbf K_{p+2}}_{II'\mathbf K_p'})
X_{\mathbf K_{p+2}J} Y^{\mathbf K_{p}'}_{J'}+\label{E:linearordpPhi}\\
&&\hp{\qquad(D^{JJ'}r^\alpha_\gamma{}_{JJ'}{}^{\mathbf K_{p+2}}_{\mathbf K_p'})
X_{\mathbf K_{p+2}I} Y^{\mathbf K_{p}'}_{I'}-2}
(\Pder{\gamma}{\mathbf K_{p+1}}{\mathbf K_{p-1}'}\tilde v^\alpha_{JJ'})
X_{\mathbf K_{p+1}I}^{J} Y^{\mathbf K_{p-1}'J'}_{I'}= 0,\nonumber
\end{eqnarray}
where we have factored out $\mbox{\rm w}^\gamma$.
First multiply equation \eqref{E:linearordpPhi}
by $X^I$ and sum over $I$ and then multiply
equation \eqref{E:linearordpPhi}
by $Y^{I'}$ and sum over $I'$ to get the equations
\[
D_{(J'}^{(J}r^{|\alpha|I\mathbf K_{p+2})}_{|\beta I'|\mathbf K_p')}=0,
\qquad
D_{(J'}^{(J}r^{|\alpha I|\mathbf K_{p+2})}_{|\beta|I'\mathbf K_p')}=0.
\]
Thus 
\begin{equation}
\label{E:Killingspinor}
D^{(J}_{(J'}s_{|\beta|}^{|\alpha|}{}_
{\mathbf K_{p+1}')}^{\mathbf K_{p+1})} = 0,\qquad
D^{(J}_{(J'}t_{|\beta|}^{|\alpha|}{}_
{\mathbf K_{p-1}')}^{\mathbf K_{p+3})}=0.
\end{equation}
It is easy to see that equations 
\eqref{E:Killingspinor} imply that 
the coefficients 
$s^\alpha_\beta{}^{\mathbf K_{p+1}'}_{\mathbf K_{p+1}}=
s^\alpha_\beta{}^{\mathbf K_{p+1}'}_{\mathbf K_{p+1}}(x^j)$,
$t^\alpha_\beta{}^{\mathbf K_{p-1}'}_{\mathbf K_{p+3}}=
t^\alpha_\beta{}^{\mathbf K_{p-1}'}_{\mathbf K_{p+3}}(x^j)$
are, in fact, functions of $x^j$ only, and hence 
that they are Killing spinors of type 
$(p,p)$ and $(p+3,p-1)$, respectively.

By virtue of \eqref{E:Killingspinor}
equations \eqref{E:linearordpPhi} simplify to
\begin{eqnarray}
&&\big((\Pder{\gamma}{\mathbf K_{p+1}}{\mathbf K_{p-1}'}
u^\alpha_\beta{}^{\mathbf L_{p+1}}_{\mathbf L_{p-1}'})
\Pvar{L}{\beta}{p+1}{p-1}{}_{JJ'}+
(\Pder{\gamma}{\mathbf K_{p+1}}{\mathbf K_{p-1}'}
\overline u^\alpha_\beta{}^{\mathbf L_{p+1}'}_{\mathbf L_{p-1}})
\Pvarbar{L}{\beta}{p+1}{p-1}{}_{JJ'}-\nonumber\\
&&\qquad
D_{JJ'}u^\alpha_\gamma{}^{\mathbf K_{p+1}}_{\mathbf K_{p-1}'}-
w^\alpha_\gamma{}_{JJ'\mathbf K_{p-1}'}^{\mathbf K_{p+1}}
+\Pder{\gamma}{\mathbf K_{p+1}}{\mathbf K_{p-1}'}\tilde v^\alpha_{JJ'})
X_{\mathbf K_{p+1}}^{J} Y^{\mathbf K_{p-1}'J'} = 0,
\label{E:ueqnfinal}
\end{eqnarray}
where 
$w^\alpha_\beta{}^{\mathbf K_{p+2}}_{\mathbf K_{p}'}$ 
is as in \eqref{E:wdefn}.
The above equations in turn imply that
\begin{equation}
\label{E:uintcond}
\Pder{\gamma}{\mathbf K_{p+1}}{\mathbf K_{p-1}'}
u^\alpha_\beta{}^{\mathbf L_{p+1}}_{\mathbf L_{p-1}'}=
\Pder{\beta}{\mathbf L_{p+1}}{\mathbf L_{p-1}'}
u^\alpha_\gamma{}^{\mathbf K_{p+1}}_{\mathbf K_{p-1}'},\quad
\Pder{\gamma}{\mathbf K_{p+1}}{\mathbf K_{p-1}'}
\overline u^\alpha_\beta{}^{\mathbf L_{p+1}'}_{\mathbf L_{p-1}}=
\Pderbar{\beta}{\mathbf L_{p+1}'}{\mathbf L_{p-1}}
u^\alpha_\gamma{}^{\mathbf K_{p+1}}_{\mathbf K_{p-1}'}.
\end{equation}

Let 
\[
\mathcal X^\alpha = \int_0^1 
u^\alpha_\beta{}^{\mathbf K_{p+1}}_{\mathbf K_{p-1}'}
(x^j,\Phi^{[p-2]},t\partial^{p-1}\Phi)
\Pvar{K}{\beta}{{p+1}}{{p-1}}dt
+ c.c.
\]
By \eqref{E:ordpQ2}, \eqref{E:uintcond}, 
we can subtract the generalized gauge 
symmetry with components 
$\nabla_{II'}\mathcal X^\alpha$ from $\wQ$ 
to get an equivalent symmetry $Q$ 
with components $Q_{II'}^\alpha$ 
given by
\begin{align}
Q_{II'}^\alpha =  
s^\alpha_\beta{}_{I'\mathbf K_p'}^{\mathbf K_{p+1}}
\Phi^\beta{}_{I\mathbf K_{p+1}}^{\mathbf K_{p}'}+
&t^\alpha_\beta{}_{I\mathbf K_{p-1}'}^{\mathbf K_{p+2}}
\Phi^\beta{}_{I'\mathbf K_{p+2}}^{\mathbf K_{p-1}'} +c.c.+\nonumber\\
&q^\alpha_\beta(x^j,\Phi^{[p-1]})a^\beta_{II'} +
\hat v^\alpha_{II'}(x^j,\Phi^{[p-1]}).\label{E:ordpQ3}
\end{align}
Substitute $Q$ in the determining equations and apply
the operator $B^{p+1}_{\rwseven,X,Y}$ to the resulting expression. 
Steps analogous to those leading to \eqref{E:ueqnfinal} 
now yield the equation
$$
(w^\alpha_\gamma{}_{JJ'\mathbf K_{p-1}'}^{\mathbf K_{p+1}}
-\Pder{\gamma}{\mathbf K_{p+1}}{\mathbf K_{p-1}'}\hat v^\alpha_{JJ'})
X_{\mathbf K_{p+1}}^{J} Y^{\mathbf K_{p-1}'J'} = 0.
$$
This together with its complex conjugate equation imply that
\[
\Pder{\beta}{(\mathbf K_{p+1}}{(\mathbf K_{p-1}'}
\hat v^{|\alpha|}{}^{K_{p+2})}_{K_p')} =
w^{\alpha}_{\beta}{}_{\mathbf K_{p}'}^{\mathbf K_{p+2}},\qquad
\Pderbar{\beta}{(\mathbf K_{p+1}'}{(\mathbf K_{p-1}}
\hat v^{|\alpha|}{}^{K_{p+2}')}_{K_p)} = 
\overline  w^{\alpha}_{\beta}{}^{\mathbf K_{p+2}'}_{\mathbf K_{p}}.
\]

Hence if we write
\[
\hat v^\alpha_{II'} =  v^\alpha_{II'} +
w^\alpha_\beta{}_{II'\mathbf K_{p-1}'}^{\mathbf K_{p+1}}
\Pvar{K}{\beta}{p+1}{p-1}+
\overline w^\alpha_\beta{}_{II'\mathbf K_{p-1}}^{\mathbf K_{p+1}'}
\Pvarbar{K}{\beta}{p+1}{p-1},
\]
we have that 
\begin{eqnarray}
&&Q_{II'}^\alpha =  
s^\alpha_\beta{}_{I'\mathbf K_p'}^{\mathbf K_{p+1}}(x^j)
\Phi^\beta{}_{I\mathbf K_{p+1}}^{\mathbf K_{p}'}+
t^\alpha_\beta{}_{I\mathbf K_{p-1}'}^{\mathbf K_{p+2}}(x^j)
\Phi^\beta{}_{I'\mathbf K_{p+2}}^{\mathbf K_{p-1}'}+\nonumber\\
&&\quad
w^\alpha_\beta{}^{\mathbf K_{p+1}}_{II'\mathbf K_{p-1}'}(x^j)
\Pvar{K}{\beta}{p+1}{p-1}+
c.c.+
q^\alpha_\beta(x^j,\Phi^{[p-1]})a^\beta_{II'}+
v^\alpha_{II'}(x^j,\Phi^{[p-1]}),\nonumber
\end{eqnarray}
where
\[\Pder{\beta}{(\mathbf K_{p+1}}{(\mathbf K_{p-1}'}
v^{|\alpha|}{}^{K_{p+2})}_{K_p')}=0,\qquad
\Pderbar{\beta}{(\mathbf K_{p+1}'}{(\mathbf K_{p-1}}
v^{|\alpha|}{}^{K_{p+2}')}_{K_p)} = 0.
\]
Thus, in particular, equations 
\eqref{E:vsymmder} hold.

We still need to show that the 
coefficient functions
$s^\alpha_\beta{}_{\mathbf K_{p+1}'}^{\mathbf K_{p+1}}$,
$t^\alpha_\beta{}_{\mathbf K_{p-1}'}^{\mathbf K_{p+3}}$
satisfy equation \eqref{E:stadjcomm} 
and that $q^\alpha_\beta$ does not 
depend on the variables $\partial^{p-1}\Phi$.
For this we first use equations 
\eqref{E:excomputation1}, 
\eqref{E:excomputation2} 
together with 
\eqref{E:rsymmetrization}
to conclude that terms 
involving  the variable 
$a^\gamma{}^{JJ'}$ in the 
equations obtained by applying 
the operator $B^{p+1}_{\rwseven,X,Y}$ 
to the determining equations for $Q$ 
yield the equation
\begin{eqnarray}
&&(\Pder{\delta}{\mathbf K_{p+1}}
{\mathbf K_{p-1}'}q^\alpha_\gamma)
X_{\mathbf K_{p+1}IJ}
Y^{\mathbf K_{p-1}'}_{I'J'}-
(c^\alpha_{\beta\gamma}
r^\beta_{\delta}{}^{\hp{JJ'}\mathbf K_{p+2}}_{JJ'\mathbf K_p'}X_{I}Y_{I'}+
r^\alpha_\beta{}^{\hp{JJ'}\mathbf K_{p+2}}_{JJ'\mathbf K_{p}'}
c^\beta_{\gamma\delta}X_{I}Y_{I'}-\nonumber\\
&&\qquad\qquad2c^\alpha_{\beta\gamma}
r^\beta_\delta{}^{\hp{II'}\mathbf K_{p+2}}_{II'\mathbf K_p'}X_{J}Y_{J'}-
2r^\alpha_\beta{}^{\hp{II'}\mathbf K_{p+2}}_{II'\mathbf K_{p}'}
c^\beta_{\gamma\delta}X_JY_{J'})
X_{\mathbf K_{p+2}}Y^{\mathbf K_p'} = 0.
\label{E:ordp1Phi}
\end{eqnarray}
First multiply \eqref{E:ordp1Phi} 
by $X^I$ and sum over $I$ and
then multiply \eqref{E:ordp1Phi} 
by $Y^{I'}$ and sum over $I'$ to
derive the equations
\[
c^\alpha_{\beta\gamma}r^\beta_\delta
{}^{(I}_{I'}{}^{\mathbf K_{p+2})}_{\mathbf K_p'}+
r^\alpha_\beta{}^{(I}_{I'}{}^{\mathbf K_{p+2})}_{\mathbf K_{p}'}
c^\beta_{\gamma\delta} =0,
\qquad
c^\alpha_{\beta\gamma}
r^\beta_\delta{}^{I\mathbf K_{p+2}}_{(I'\mathbf K_p')}+
r^\alpha_\beta{}^{I\mathbf K_{p+2}}_{(I'\mathbf K_{p}')}
c^\beta_{\gamma\delta}=0,
\]
that is,
\begin{equation}
\label{E:stadjcomm2}
c^\alpha_{\beta\gamma}s^\beta_\delta{}^
{\mathbf K_{p+1}}_{\mathbf K_{p+1}'}+
s^\alpha_\beta{}^{\mathbf K_{p+1}}_{\mathbf K_{p+1}'}
c^\beta_{\gamma\delta}=0,
\qquad
c^\alpha_{\beta\gamma}
t^\beta_\delta{}^{\mathbf K_{p+3}}_{\mathbf K_{p-1}'}+
t^\alpha_\beta{}^{\mathbf K_{p+3}}_{\mathbf K_{p-1}'}
c^\beta_{\gamma\delta}=0.
\end{equation}
Hence \eqref{E:stadjcomm} holds. Recall that by 
\eqref{E:ordpQ3}, the contraction
$r^\alpha_\beta{}_{K_{p+2}}^{K_p'}{}_
{\mathbf K_{p-1}'K_p'}^{\mathbf K_{p+1}K_{p+2}}$
vanishes.
Thus with \eqref{E:stadjcomm2}, equation 
\eqref{E:ordp1Phi} and its complex conjugate 
equation furthermore imply that
$$q^\alpha_\beta=q^\alpha_\beta(x^j,\Phi^{[p-2]}).$$
This concludes the proof of Lemma 
\ref{L:ordpPhired}.
\end{proof}

Next let $\wQ$ be a symmetry with 
components $\wQ_{II'}^\alpha$ of the form 
\begin{eqnarray}
&&\wQ_{II'}^\alpha =  
s^\alpha_\beta{}_{I'\mathbf K_p'}^{\mathbf K_{p+1}}(x^j)
\Phi^\beta{}_{I\mathbf K_{p+1}}^{\mathbf K_{p}'}+
t^\alpha_\beta{}_{I\mathbf K_{p-1}'}^{\mathbf K_{p+2}}(x^j)
\Phi^\beta{}_{I'\mathbf K_{p+2}}^{\mathbf K_{p-1}'}+
w^\alpha_\beta{}^{\mathbf K_{p+1}}_{II'\mathbf K_{p-1}'}(x^j)
\Pvar{K}{\beta}{p+1}{p-1}+ \nonumber\\ 
&&\qquad\qquad\qquad\qquad
c.c.+\tilde q^\alpha_\beta(x^j,\Phi^{[p-2]})a^\beta_{II'} +
v^\alpha_{II'}(x^j,\Phi^{[p-1]}),\quad p\geq1,
\label{E:ordpQ4}
\end{eqnarray} 
where the functions 
$s^\alpha_\beta{}_{\mathbf K_{p+1}'}^{\mathbf K_{p+1}}$,
$t^\alpha_\beta{}_{\mathbf K_{p-1}'}^{\mathbf K_{p+3}}$,
$w^\alpha_\beta{}^{\mathbf K_{p+2}}_{\mathbf K_{p}'}$,
$v^\alpha_{II'}$ satisfy \eqref{E:stadjcomm}, 
\eqref{E:wdefn}, \eqref{E:vsymmder}.
We now analyze terms involving $\partial^p\Phi$
in the determining equations for $\wQ$.
\begin{lemma}\label{L:ordp1Phired}
Let $\wQ$ be a symmetry of the Yang-Mills fields
with components $\wQ^\alpha_{II'}$ as in 
\eqref{E:ordpQ4}. Then $\wQ$ is equivalent to 
a symmetry $Q$ with components $Q_{II'}^\alpha$
given by
\begin{align}
Q_{II'}^\alpha =\,
&s^\alpha_\beta{}_{I'\mathbf K_p'}^{\mathbf K_{p+1}}(x^j)
\Phi^\beta{}_{I\mathbf K_{p+1}}^{\mathbf K_{p}'}+
t^\alpha_\beta{}_{I\mathbf K_{p-1}'}^{\mathbf K_{p+2}}(x^j)
\Phi^\beta{}_{I'\mathbf K_{p+2}}^{\mathbf K_{p-1}'}+
\nonumber\\
&\hat s^\alpha_\beta{}_{I'\mathbf K_{p-1}'}^{\mathbf K_{p}}(x^j)
\Phi^\beta{}_{I\mathbf K_{p}}^{\mathbf K_{p-1}'}+
\hat t^\alpha_\beta{}_{I\mathbf K_{p-2}'}^{\mathbf K_{p+1}}(x^j)
\Phi^\beta{}_{I'\mathbf K_{p+1}}^{\mathbf K_{p-2}'}+
\label{E:ordp1Q1}\\
&
w^\alpha_\beta{}^{\mathbf K_{p+1}}_{II' \mathbf K_{p-1}'}(x^j)
\Pvar{K}{\beta}{p+1}{p-1}+
c.c. + q^\alpha_\beta(x^j,\Phi^{[p-2]})a^\beta_{II'} +
z^\alpha_{II'}(x^j,\Phi^{[p-2]}),\nonumber
\end{align}
where the functions 
$\hat s^\alpha_\beta{}_{\mathbf K_{p}'}^{\mathbf K_{p}}$,
$\hat t^\alpha_\beta{}_{\mathbf K_{p-2}'}^{\mathbf K_{p+2}}$
are symmetric in their spinorial indices.
\end{lemma}
\begin{proof} 
Define 
$r^\alpha_\beta{}_{II'}{}^{\mathbf K_{p+2}}_{\mathbf K_p'}$
by
\[
r^\alpha_\beta{}_{II'}{}^{\mathbf K_{p+2}}_{\mathbf K_p'}
\Pvar{K}{\beta}{p+2}{p} = 
s^\alpha_\beta{}_{I'\mathbf K_p'}^{\mathbf K_{p+1}}
\Phi^\beta{}_{I\mathbf K_{p+1}}^{\mathbf K_{p}'}+
t^\alpha_\beta{}_{I\mathbf K_{p-1}'}^{\mathbf K_{p+2}}
\Phi^\beta{}_{I'\mathbf K_{p+2}}^{\mathbf K_{p-1}'}.
\]
By the proof of Lemma \ref{L:ordpPhired}, 
conditions \eqref{E:stadjcomm}, 
\eqref{E:wdefn}, \eqref{E:vsymmder}
guarantee that the highest order
terms in the determining equations for 
$\wQ$ are $\partial^p\Phi$. 
Note that due to condition \eqref{E:stadjcomm}, 
we can apply equation \eqref{E:productrulesimple}
in Proposition \ref{P:productrule} in computing 
covariant derivatives of the term 
$r^\alpha_\beta{}_{II'}{}^{\mathbf K_{p+2}}_{\mathbf K_p'}
\Pvar{K}{\beta}{p+2}{p}$.
Thus, when substituted in the determining equations 
\eqref{E:YMsymmspinor}, it yields the expression
\begin{eqnarray}
&&r^\alpha_\beta{}^{\hp{JJ'}\mathbf K_{p+2}}_{JJ'\mathbf K_p'}
\nabla^{JJ'}\nabla_{II'}\Pvar{K}{\beta}{p+2}{p}-
r^\alpha_\beta{}^{\hp{II'}\mathbf K_{p+2}}_{II'\mathbf K_p'}
\nabla^{JJ'}\nabla_{JJ'}\Pvar{K}{\beta}{p+2}{p}+\nonumber\\
&&(\partial_{II'}r^\alpha_\beta{}^{JJ'\mathbf K_{p+2}}_{\hp{JJ'}\mathbf K_p'}-
2\partial^{JJ'}r^\alpha_\beta{}^{\hp{II'}\mathbf K_{p+2}}_{II'\mathbf K_p'})
\nabla_{JJ'}\Pvar{K}{\beta}{p+2}{p}+
(\partial^{JJ'}r^\alpha_\beta{}^{\hp{JJ'}\mathbf K_{p+2}}_{JJ'\mathbf K_p'})
\nabla_{II'}\Pvar{K}{\beta}{p+2}{p}+\nonumber\\
&&\qquad\qquad(\partial^{JJ'}\partial_{II'}
r^\alpha_\beta{}^{\hp{JJ'}\mathbf K_{p+2}}_{JJ'\mathbf K_p'}-
\partial^{JJ'}\partial_{JJ'}
r^\alpha_\beta{}^{\hp{II'}\mathbf K_{p+2}}_{II'\mathbf K_p'})
\Pvar{K}{\beta}{p+2}{p}-\label{E:rsymmeqn}\\
&&\qquad\qquad\qquad\qquad\qquad\qquad
c^\alpha_{\beta\gamma}(
\Phi^\beta{}_I^J
r^\gamma_\delta{}
^{\hp{JI'}\mathbf K_{p+2}}_{JI'\mathbf K_{p'}}+
\overline \Phi^\beta{}_{I'}^{J'} 
r^\gamma_\delta{}
^{\hp{IJ'}\mathbf K_{p+2}}_{IJ'\mathbf K_{p'}})\Pvar{K}{\delta}{p+2}{p}.
\nonumber
\end{eqnarray}
Now by \eqref{E:derPhisymmvar}, it is clear 
that \eqref{E:rsymmeqn} gives rise to terms of 
order $p+1$ that are either linear expressions in
$\partial^p\Phi$, $\partial^p\overline\Phi$
or in the products of either $\Phi$, $\overline\Phi$ 
with 
$\partial^p\Phi$ or $\partial^p\overline\Phi$
with coefficients that are functions of $x^j$ only.
Thus, since $p\geq1$, the only terms in the 
determining equations for $\wQ$
involving quadratic expressions in 
$\partial^p\Phi$, $\partial^p\overline\Phi$
arise from the function $v^\alpha_{II'}$. Hence,
steps identical to those leading to equation 
\eqref{E:scndderpr} in the proof of Lemma 
\ref{L:ordpPhired} show that 
$v^\alpha_{II'}$ must be linear in
the variables $\partial^{p-1}\Phi$ with 
coefficients that are functions of the
variables $x^j$, 
$\Phi^{[p-2]}$. In particular, 
if $p=1$, it follows that equation 
\eqref{E:ordp1Q1} holds with 
$\hat t^\alpha_\beta{}_{\mathbf K_{p-2}'}^{\mathbf K_{p+2}}=0$.

Now assume that $p\geq2$. Write 
\begin{eqnarray}
&&w^\alpha_\beta{}^{\mathbf K_{p+1}}_{II'\mathbf K_{p-1}'}(x^j)
\Pvar{K}{\beta}{p+1}{p-1}+
\overline w^\alpha_\beta{}^{\mathbf K'_{p+1}}_{II'\mathbf K_{p-1}}(x^j)
\Pvarbar{K}{\beta}{p+1}{p-1}+
v^\alpha_{II'}(x^j,\Phi^{[p-1]}) =\nonumber\\
&&
\hat r^\alpha_\beta{}
^{\hp{II'}\mathbf K_{p+1}}_{II'\mathbf K_{p-1}'} (x^j,\Phi^{[p-2]})
\Pvar{K}{\beta}{p+1}{p-1}  + 
\overline{\hat r}^\alpha_\beta{}
^{\hp{II'}\mathbf K_{p+1}'}_{II'\mathbf K_{p-1}} (x^j,\Phi^{[p-2]})
\Pvarbar{K}{\beta}{p+1}{p-1}+ 
\hat z^\alpha_{II'}(x^j,\Phi^{[p-2]}),\nonumber
\end{eqnarray}
so that
$\hat r^\alpha_\beta{}^{(I}_{(I'}{}^{\mathbf K_{p+1})}_{\mathbf K_{p-1}')}=
w^\alpha_\beta{}^{I\mathbf K_{p+1}}_{I'\mathbf K_{p-1}'}$. 
By virtue of Proposition \ref{P:dersymmvar} we have that
\begin{align}
\nabla_{K_{p+4}}^{K_{p+2}'}\nabla_{K_{p+3}}^{K_{p+1}'}
\Pvar{K}{\alpha}{p+2}{p}&=
\Pvar{K}{\alpha}{p+4}{p+2} +
G^\alpha_\beta{}_{K_{p+4}K_{p+3}\mathbf K_{p+2}}^
{K_{p+2}'K_{p+1}'\mathbf K_{p}'},
{}^{\mathbf L_{p+2}}_{\mathbf L_p'}
\Pvar{L}{\beta}{p+2}{p}+\nonumber\\
&\qquad H^\alpha_\beta{}_{K_{p+4}K_{p+3}\mathbf K_{p+2}}^
{K_{p+2}'K_{p+1}'\mathbf K_{p}'},
{}^{\mathbf L_{p+2}'}_{\mathbf L_p}
\Pvarbar{L}{\beta}{p+2}{p}+
\widetilde\Psi^\alpha{}_{\mathbf K_{p+4}}^{\mathbf K_{p+2}'},
\nonumber
\end{align}
where $G^\alpha_\beta{}_{K_{p+4}K_{p+3}\mathbf K_{p+2}}^
{K_{p+2}'K_{p+1}'\mathbf K_{p}'},{}^{\mathbf L_{p+2}}_{\mathbf L_p'}$, 
$H^\alpha_\beta{}_{K_{p+4}K_{p+3}\mathbf K_{p+2}}^
{K_{p+2}'K_{p+1}'\mathbf K_{p}'},{}^{\mathbf L_{p+2}'}_{\mathbf L_p}$ 
are constant coefficient linear expressions 
in the field variables $\Phi^\alpha_{IJ}$, 
$\overline\Phi^\alpha_{I'J'}$, and 
$\widetilde\Psi^\alpha{}_{\mathbf K_{p+4}}^{\mathbf K_{p+2}'}$, 
when restricted to the solution manifold $\mathcal R^p$, 
are functions of the variables $\Phi^{[p-1]}$ only.
Now apply the operator $B^p_{\rvseven,X,Y}$ to the 
determining equations for $\wQ$ 
and factor out $\rv^\gamma$ to see that 
on $\mathcal R^p$,

\begin{eqnarray}
&&(r^\alpha_\beta{}^{\hp{JJ'}\mathbf L_{p+2}}_{JJ'\mathbf L_p'}
G^\beta_\gamma{}_{\hp{JJ'}II'\mathbf L_{p+2}}^{JJ'\hp{II'}\mathbf L_p'},{}
_{\mathbf K_{p}'}^{\mathbf K_{p+2}}-
r^\alpha_\beta{}^{\hp{II'}\mathbf L_{p+2}}_{II'\mathbf L_p'}
G^\beta_\gamma{}^{JJ'}{}_{JJ'\mathbf L_{p+2}}^{\hp{JJ'}\mathbf L_p'},{}_
{\mathbf K_{p}'}^{\mathbf K_{p+2}}+\nonumber\\
&&\overline r^\alpha_\beta{}^{\hp{JJ'}\mathbf L_{p+2}'}_{JJ'\mathbf L_p}
\overline H^\beta_\gamma{}^{JJ'}{}_{II'\mathbf L_{p+2}'}^{\hp{II'}\mathbf L_p},{}
_{\mathbf K_{p}'}^{\mathbf K_{p+2}}-
\overline r^\alpha_\beta{}^{\hp{II'}\mathbf L_{p+2}'}_{II'\mathbf L_p}
\overline H^\beta_\gamma{}^{JJ'}{}_{JJ'\mathbf L_{p+2}'}^{\hp{JJ'}\mathbf L_p},{}_
{\mathbf K_{p}'}^{\mathbf K_{p+2}}+\nonumber\\
&&\partial^{JJ'}\partial_{II'} 
r^\alpha_\gamma{}_{JJ'\mathbf K_{p}'}^{\hp{JJ'}\mathbf K_{p+2}}-
\partial^{JJ'}\partial_{JJ'} 
r^\alpha_\gamma{}_{II'\mathbf K_{p}'}^{\hp{II'}\mathbf K_{p+2}})
X_{\mathbf K_{p+2}}Y^{\mathbf K_p'}+\label{E:ordpPhiterms}
\end{eqnarray}
\begin{eqnarray}
&&\big((\Pder{\gamma}{\mathbf K_{p}}{\mathbf K_{p-2}'}
\hat r^\alpha_\beta{}^{\hp{JJ'}\mathbf L_{p+1}}_{JJ'\mathbf L_{p-1}'})
\Pvar{L}{\beta}{p+1}{p-1}+
(\Pder{\gamma}{\mathbf K_{p}}{\mathbf K_{p-2}'}
\overline {\hat r}^\alpha_\beta{}^{\hp{JJ'}\mathbf L_{p+1}'}_{JJ'\mathbf L_{p-1}}\big)
\Pvarbar{L}{\beta}{p+1}{p-1})X_{\mathbf K_pI}^{J} Y^{\mathbf K_{p-2}'J'}_{I'}+
\nonumber\\
&&\big((D^{JJ'}\hat r^\alpha_\gamma{}^{\hp{JJ'}\mathbf K_{p+1}}_{JJ'\mathbf K_{p-1}'})
X_{I} Y_{I'}-
(2D^{JJ'}\hat r^\alpha_\gamma{}^{\hp{II'}\mathbf K_{p+1}}_{II'\mathbf K_{p-1}'}-
\partial_{II'}w_\gamma^\alpha{}^{JJ'}{}^{\mathbf K_{p+1}}_{\mathbf K_{p-1}'})X_JY_{J'}\big)
X_{\mathbf K_{p+1}} Y^{\mathbf K_{p-1}'}+\nonumber\\
&&(\Pder{\gamma}{\mathbf K_{p}}{\mathbf K_{p-2}'} \hat z^\alpha_{JJ'})
X_{\mathbf K_pI}^{J} Y^{\mathbf K_{p-2}'J'}_{I'}+
a^\beta_{LL'} S^{ \alpha }_{\beta\gamma}{}^{LL'}_{II'}{}
^{\mathbf K_{p+2}}_{\mathbf K_p'}
X_{\mathbf K_{p+2}}Y^{\mathbf K_p'}-\nonumber\\
&&c^\alpha_{\beta\delta}(\Phi^\beta{}_I^J
r^\delta_\gamma{}
^{\hp{JI'}\mathbf K_{p+2}}_{JI'\mathbf K_{p'}}+
\overline \Phi^\beta{}_{I'}^{J'} r^\delta_\gamma{}
^{\hp{IJ'}\mathbf K_{p+2}}_{IJ'\mathbf K_{p'}})
X_{\mathbf K_{p+2}}Y^{\mathbf K_p'}=0,
\nonumber
\end{eqnarray}
where the terms
$S^{\alpha}_{\beta\gamma}{}^{LL'}_{II'}{}^{\mathbf K_{p+2}}_{\mathbf K_p'}$ 
are determined by the equation
\begin{align}
&a^\beta_{LL'}S^\alpha_{\beta\gamma}{}^{LL'}_
{II'}{}^{\mathbf K_{p+2}}_{\mathbf K_p'}
\Phi^\gamma{}^{\mathbf K_p'}_{\mathbf K_{p+2}}=\nonumber\\
-&(c^\alpha_{\beta\gamma}
\hat r^\beta_\delta{}^{\hp{JJ'}\mathbf K_{p+1}}_{JJ'\mathbf K_{p-1}'}+
\hat r^\alpha_\beta{}^{\hp{JJ'}\mathbf K_{p+1}}_
{JJ'\mathbf K_{p-1}'}c^\beta_{\gamma\delta})
(a^\gamma{}^{JJ'}\Phi^\delta{}^{\mathbf K_{p-1}'}_{\mathbf K_{p+1}II'}+
a^\gamma_{II'}\Phi^\delta{}^{\mathbf K_{p-1}'JJ'}_{\mathbf K_{p+1}})+
\nonumber\\
2&(c^\alpha_{\beta\gamma}
\hat r^\beta_\delta{}^{\hp{II'}\mathbf K_{p+1}}_{II'\mathbf K_{p-1}'}+
\hat r^\alpha_\beta{}^{\hp{II'}\mathbf K_{p+1}}_
{II'\mathbf K_{p-1}'}c^\beta_{\gamma\delta})
a^\gamma{}^{JJ'}\Phi^\delta{}^{\mathbf K_{p-1}'}_{\mathbf K_{p+1}JJ'}+
(\partial_{\Phi^\gamma}{}_{\mathbf K_{p-2}'}^{\mathbf K_ p}q^\alpha_\beta) 
a^\beta_{JJ'}\Phi^\gamma{}_{\mathbf K_p}^{\mathbf K_{p-2}'}{}_{II'}^{JJ'}.
\nonumber
\end{align}

Our next goal is to show that equation 
\eqref{E:ordpPhiterms} forces the symmetrizations
$\hat r^\alpha_\beta{}_{I'\mathbf K_{p-1}'}^{(I\mathbf K_{p+1})}$,
$\hat r^\alpha_\beta{}_{(I'\mathbf K_{p-1}')}^{I\mathbf K_{p+1}}$
to be functions of $x^j$ only. Suppose on the contrary that
$\hat r^\alpha_\beta{}_{I'}^{(I}{}_{\mathbf K_{p-1}'}^{\mathbf K_{p+1})}$,
$\hat r^\alpha_\beta{}_{(I'\mathbf K_{p-1}')}^{I\mathbf K_{p+1}}$
depend on the variables 
$\Phi^{[q]}$ for some $0\leq q\leq p-2$.
Multiply \eqref{E:ordpPhiterms} by $X^I$ 
and sum over $I$. In the resulting equations 
the only terms involving the variables
$\partial^{q+1}\Phi$ arise from the total derivative term
$D^{JJ'}\hat r^\alpha_\beta{}_{II'\mathbf K_{p-1}'}^{\hp{II'}\mathbf K_{p+1}}$.
Hence upon an application of the operator 
$B^{q+1}_{\rwseven,Z,W}$ we get the equation
\[
(\Pder{\delta}{\mathbf L_{q+2}}{\mathbf L_{q}'}
\hat r^\alpha_\gamma{}^{\hp{II'}\mathbf K_{p+1}}_{II'\mathbf K_{p-1}'})
X_{\mathbf K_{p+1}J}^I Y_{J'}^{\mathbf K_{p-1}'}
Z_{\mathbf L_{q+2}}^J W^{\mathbf L_{q}'J'} = 0,
\]
from which it follows that
\[
\Pder{\delta}{\mathbf L_{q+2}}{\mathbf L_{q}'}
\hat r^\alpha_\gamma{}^{(I}_{I'}{}^{\mathbf K_{p+1})}_{\mathbf K_{p-1}'}=0.
\]
One can similarly show that
\[
\Pderbar{\delta}{\mathbf L_{q+2}'}{\mathbf L_{q}}
\hat r^\alpha_\gamma{}^{(I}_{I'}{}^{\mathbf K_{p+1})}_{\mathbf K_{p-1}'}=0,\quad
\Pder{\delta}{\mathbf L_{q+2}}{\mathbf L_{q}'}
\hat r^\alpha_\gamma{}^{I\mathbf K_{p+1}}_{(I'\mathbf K_{p-1}')}=0,\quad
\Pderbar{\delta}{\mathbf L_{q+2}'}{\mathbf L_{q}}
\hat r^\alpha_\gamma{}^{I\mathbf K_{p+1}}_{(I'\mathbf K_{p-1}')}=0,
\]
which contradicts our assumption. Hence we have that
\begin{eqnarray}
&&\hat r^\alpha_\beta{}^{\hp{II'}\mathbf K_{p+1}}_{II'\mathbf K_{p-1}'}
\Pvar{K}{\beta}{p+1}{p-1}=
\hat s^\alpha_\beta{}_{I'\mathbf K_{p-1}'}^{\mathbf K_{p}}(x^j)
\Phi^\beta{}_{I\mathbf K_{p}}^{\mathbf K_{p-1}'}+
\hat t^\alpha_\beta{}_{I\mathbf K_{p-2}'}^{\mathbf K_{p+1}}(x^j)
\Phi^\beta{}_{I'\mathbf K_{p+1}}^{\mathbf K_{p-2}'}+\nonumber\\
&&\qquad\qquad
w^\alpha_\beta{}^{\mathbf K_{p+1}}_{II' \mathbf K_{p-1}'}(x^j)
\Pvar{K}{\beta}{p+1}{p-1}+
\hat u^\alpha_\beta{}_{\mathbf K_{p-2}'}^{\mathbf K_{p}}(x^j,\Phi^{[p-2]})
\Phi^\beta{}^{\mathbf K_{p-2}'}_{II'\mathbf K_p},\label{E:ordp-1Qterms}
\end{eqnarray}
where $\hat s^\alpha_\beta{}_{\mathbf K_{p}'}^{\mathbf K_{p}}$,
$\hat t^\alpha_\beta{}_{\mathbf K_{p-2}'}^{\mathbf K_{p+2}}$
are symmetric in their spinorial indices.
Now substitute \eqref{E:ordp-1Qterms} into 
\eqref{E:ordpPhiterms}. Since $p\geq2$, the only
terms in the resulting equations involving the variables 
$\partial^{p-1}\Phi$ arise from the terms on lines 4 and 5 
in \eqref{E:ordpPhiterms}. Specifically, these terms 
yield the equation
\begin{eqnarray}
&&\big((\Pder{\gamma}{\mathbf K_{p}}{\mathbf K_{p-2}'}
\hat u^\alpha_\beta{}^{\mathbf L_{p}}_{\mathbf L_{p-2}'})
\Phi^{\beta}{}_{\hp{JJ'}\mathbf L_p}^{JJ'\mathbf L_{p-2}'}+
(\Pder{\gamma}{\mathbf K_{p}}{\mathbf K_{p-2}'}
\overline {\hat u}^\alpha_\beta{}^{\mathbf L_{p}'}_{\mathbf L_{p-2}})
\overline\Phi^{\beta}{}_{\hp{JJ'}\mathbf L_p'}^{JJ'\mathbf L_{p-2}}-\nonumber\\
&&\qquad(\Pder{\beta}{\mathbf L_p}{\mathbf L_{p-2}'}
\hat u^\alpha_\gamma{}^{\mathbf K_{p}}_{\mathbf K_{p-2}'})
\Phi^{\beta}{}^{JJ' \mathbf L_{p-2}'}_{\hp{JJ'}\mathbf L_p}-
(\Pderbar{\beta}{\mathbf L_p'}{\mathbf L_{p-2}}
\hat u^\alpha_\gamma{}^{\mathbf K_{p}}_{\mathbf K_{p-2}'})
\overline{\Phi}^{\beta}{}^{JJ' \mathbf L_{p-2}}_{\hp{JJ'}\mathbf L_p'}\big)
X_{IJ\mathbf K_{p}} Y^{\mathbf K_{p-2}'}_{I'J'}=0,\nonumber
\end{eqnarray}
which implies that
\[
\Pder{\gamma}{\mathbf L_{p}}{\mathbf L_{p-2}'}
\hat u^\alpha_\beta{}^{\mathbf K_{p}}_{\mathbf K_{p-2}'}=
\Pder{\beta}{\mathbf K_{p}}{\mathbf K_{p-2}'}
\hat u^\alpha_\gamma{}^{\mathbf L_{p}}_{\mathbf L_{p-2}'},\quad
\Pderbar{\beta}{\mathbf L_{p}'}{\mathbf L_{p-2}}
\hat u^\alpha_\gamma{}^{\mathbf K_{p}}_{\mathbf K_{p-2}'}=
\Pder{\gamma}{\mathbf K_{p}}{\mathbf K_{p-2}'}
\overline {\hat u}^\alpha_\beta{}^{\mathbf L_{p}'}_{\mathbf L_{p-2}}.
\]
Thus we can subtract a gauge symmetry from 
$\wQ$ to obtain a symmetry of the 
required form \eqref{E:ordp1Q1}. 
This concludes the proof of 
Lemma \ref{L:ordp1Phired}.
\end{proof}

We now come to the crucial step in the proof of 
Theorem \ref{T:maintheorem}, which consists of 
analyzing terms in the determining equations 
for a symmetry $Q$ in \eqref{E:ordp1Q1} 
involving products of the variables 
$\Phi$ and $\partial^p\Phi$. 
\begin{lemma}\label{L:PhiordpPhi}
Let $Q$ be a symmetry of the Yang-Mills equations
with components $Q^\alpha_{II'}$ as in \eqref{E:ordp1Q1} 
with $p\geq1$. Then 
\[
s^\alpha_\beta{}_{\mathbf K_{p+1}'}^{\mathbf K_{p+1}}=0,\qquad
t^\alpha_\beta{}_{\mathbf K_{p-1}'}^{\mathbf K_{p+3}}=0.
\]
\end{lemma}
\begin{proof} 
Write $\mathcal E^\alpha_{II'}(P^\beta_{JJ'})$ 
for the expression obtained
by substituting differential functions 
$P^\beta_{JJ'}$ on the left-hand side of
the determining equations \eqref{E:YMsymmspinor}.
Our goal is to analyze the equations
\begin{equation}\label{E:PhiordpPhi}
\mbox{B}^o_{\rwseven,Z}\mbox{B}^p_{\rvseven,X,Y}
\mathcal E^\alpha_{II'}(Q^\beta_{JJ'})=0,
\end{equation}
where the operators $\mbox{B}^o_{\rwseven,Z}$,
$\mbox{B}^p_{\rvseven,X,Y}$ are as in \eqref{E:Boperator}.

First note that the functions $z^\alpha_{II'}$, 
when substituted in the determining equations, 
yield the expression
\[
\mathcal E^\alpha_{II'}(z^\beta_{JJ'})=
(\partial_{\Phi^\gamma}{}^{\mathbf K_{p}}_{\mathbf K_{p-2}'}z^\alpha_{JJ'})
\Phi^{\gamma}{}_{II'\mathbf K_p}^{JJ'\mathbf K_{p-2}'}+
(\overline\partial_{\Phi^\gamma}{}^{\mathbf K_{p}'}_{\mathbf K_{p-2}}z^\alpha_{JJ'})
\overline\Phi^{\gamma}{}_{II'\mathbf K_p'}^{JJ'\mathbf K_{p-2}}+
\Psi^\alpha_{1,II'},
\] 
where $\Psi^\alpha_{1,II'}$ is of order $p$. 
Consequently,
\begin{equation}\label{E:zPhiordpPhi}
\mbox{B}^o_{\rwseven,Z}\mbox{B}^p_{\rvseven,X,Y}
\mathcal E^\alpha_{II'}(z^\beta_{JJ'})=
(\partial_{\Phi^\delta}^{LM}
\partial_{\Phi^\gamma}{}^{\mathbf K_{p}}_{\mathbf K_{p-2}'}z^\alpha_{JJ'})
Z_{LM}X_{I\mathbf K_p}^J Y^{J'\mathbf K_{p-2}'}_{I'} 
\rv^\gamma \rw^\delta.
\end{equation}
One similarly shows that
\begin{equation}\label{E:qPhiordpPhi}
\mbox{B}^o_{\rwseven,Z}\mbox{B}^p_{\rvseven,X,Y}
\mathcal E^\alpha_{II'}(q^\beta_\gamma a^\gamma_{JJ'})=
(\partial_{\Phi^\delta}^{LM}
\partial_{\Phi^\gamma}{}^
{\mathbf K_{p}}_{\mathbf K_{p-2}'}q^\alpha_\beta)a^\beta_{JJ'}
Z_{LM}X_{I\mathbf K_p}^J Y^{J'\mathbf K_{p-2}'}_{I'} 
\rv^\gamma \rw^\delta.
\end{equation}

Next note that the terms 
\[
\hat s^\alpha_\beta{}_{I'\mathbf K_{p-1}'}^{\mathbf K_{p}}
\Phi^\beta{}_{I\mathbf K_{p}}^{\mathbf K_{p-1}'},\quad
\hat t^\alpha_\beta{}_{I\mathbf K_{p-2}'}^{\mathbf K_{p+1}}
\Phi^\beta{}_{I'\mathbf K_{p+1}}^{\mathbf K_{p-2}'},\quad
w^\alpha_\beta{}^{\mathbf K_{p+1}}_{II' \mathbf K_{p-1}'}
\Pvar{K}{\beta}{p+1}{p-1}
\]
and their complex conjugate terms in \eqref{E:ordp1Q1}, 
when substituted in the determining equations
\eqref{E:YMsymmspinor}, only yield terms of order 
$p+1$ that are linear in the variables 
$\partial^p\Phi$ with coefficients that 
are functions of $x^j$. Consequently, 
for $T^\beta_{JJ'}$ any of the terms
\begin{eqnarray}
&&\hat s^\beta_\gamma{}_{J'\mathbf K_{p-1}'}^{\mathbf K_{p}}
\Phi^\gamma{}_{J\mathbf K_{p}}^{\mathbf K_{p-1}'},\quad
\overline{\hat s}^\beta_\gamma{}_{J\mathbf K_{p-1}}^{\mathbf K_{p}'}
\overline\Phi^\gamma{}_{J'\mathbf K_{p}'}^{\mathbf K_{p-1}},\quad
\hat t^\beta_\gamma{}_{J\mathbf K_{p-2}'}^{\mathbf K_{p+1}}
\Phi^\gamma{}_{J'\mathbf K_{p+1}}^{\mathbf K_{p-2}'},
\nonumber\\
&&\overline{\hat t}^\beta_\gamma{}_{J'\mathbf K_{p-2}}^{\mathbf K_{p+1}'}
\overline{\Phi}^\gamma{}_{J\mathbf K_{p+1}'}^{\mathbf K_{p-2}},\quad
w^\beta_\gamma{}^{\mathbf K_{p+1}}_{JJ' \mathbf K_{p-1}'}
\Pvar{K}{\gamma}{p+1}{p-1},\quad
\overline w^\beta_\gamma{}^{\mathbf K_{p+1}'}_{JJ' \mathbf K_{p-1}}
\Pvarbar{K}{\gamma}{p+1}{p-1},\nonumber
\end{eqnarray}
we have that
\begin{equation}
\label{E:PhiOrdp1Phi}
\mbox{B}^o_{\rwseven,Z}\mbox{B}^p_{\rvseven,X,Y}\mathcal E^\alpha_{II'}
(T^\beta_{JJ'})=0.
\end{equation}

Note that
\begin{eqnarray}
&&\mbox{B}^o_{\rwseven,Z}\mbox{B}^p_{\rvseven,X,Y}\mathcal E^\alpha_{II'}(
s^\beta_\gamma{}_{J'\mathbf K_p'}^{\mathbf K_{p+1}}                     
\Phi^\gamma{}_{J\mathbf K_{p+1}}^{\mathbf K_{p}'})=\nonumber\\
&&\qquad\qquad B^o_{\rwseven,Z}\mbox{B}^p_{\rvseven,X,Y}
(s^\alpha_\beta{}_{J'\mathbf K_p'}^{\mathbf K_{p+1}}
\nabla^{JJ'}\nabla_{II'}\Phi^\beta{}_{J\mathbf K_{p+1}}^{\mathbf K_{p}'}-
\label{E:sPhiordpPhi}\\
&&\qquad\qquad\qquad\qquad
s^\alpha_\beta{}_{I'\mathbf K_p'}^{\mathbf K_{p+1}}
\nabla^{JJ'}\nabla_{JJ'}\Phi^\beta{}_{I\mathbf K_{p+1}}^{\mathbf K_{p}'}
-s^\alpha_\beta{}^{\mathbf K_{p+1}}_{I'\mathbf K_p'}
[\Phi^J_I,\Phi^{\mathbf K_p'}_{J\mathbf K_{p+1}}]^\beta),
\nonumber\end{eqnarray}
and
\begin{eqnarray}
&&\mbox{B}^o_{\rwseven,Z}\mbox{B}^p_{\rvseven,X,Y}\mathcal E^\alpha_{II'}(
t^\beta_\gamma{}_{J\mathbf K_{p-1}'}^{\mathbf K_{p+2}}
\Phi^\gamma{}_{J'\mathbf K_{p+2}}^{\mathbf K_{p-1}'}) =\nonumber\\
&&\qquad\quad
B^o_{\rwseven,Z}\mbox{B}^p_{\rvseven,X,Y}(
t^\alpha_\beta{}_{J\mathbf K_{p-1}'}^{\mathbf K_{p+2}}
\nabla^{JJ'}\nabla_{II'}\Phi^\beta{}_{J'\mathbf K_{p+2}}^{\mathbf K_{p-1}'}-
\label{E:tPhiordpPhi}\\
&&\qquad\qquad\qquad
t^\alpha_\beta{}_{I\mathbf K_{p-1}'}^{\mathbf K_{p+2}}
\nabla^{JJ'}\nabla_{JJ'}\Phi^\beta{}_{I'\mathbf K_{p+2}}^{\mathbf K_{p-1}'}-
t^\alpha_\beta{}_{J\mathbf K_{p-1}'}^{\mathbf K_{p+2}}
[\Phi^J_I,\Phi_{I'\mathbf K_{p+2}}^{\mathbf K_{p-1}'}]^\beta).
\nonumber\end{eqnarray}
We use equation \eqref{E:covarcommutespin} to compute
\begin{align}
s^\alpha_\beta{}_{J'\mathbf K_p'}^{\mathbf K_{p+1}}
&\nabla^{JJ'}\nabla_{II'}\Phi^\beta{}_{J\mathbf K_{p+1}}^{\mathbf K_{p}'} =
s^\alpha_\beta{}_{J'\mathbf K_p'}^{\mathbf K_{p+1}}                  
\nabla_{II'}\nabla^{JJ'}\Phi^\beta{}_{J\mathbf K_{p+1}}^{\mathbf K_{p}'}+
\nonumber\\
&\qquad\qquad
s^\alpha_\beta{}_{I'\mathbf K_p'}^{\mathbf K_{p+1}}
[\Phi_{SI},\Phi_{\mathbf K_{p+1}}^{\mathbf K_p' S}]^\beta-
s^\alpha_\beta{}_{J'\mathbf K_p'}^{\mathbf K_{p+1}}
[\overline\Phi^{J'}_{I'},\Phi_{I\mathbf K_{p+1}}^{\mathbf K_p'}]^\beta.
\label{E:SndDersTerm}
\end{align}
But by equation \eqref{E:derPhisymmvar},
\begin{eqnarray}
&&s^\alpha_\beta{}_{J'\mathbf K_p'}^{\mathbf K_{p+1}}
\nabla_{II'}\nabla^{JJ'}\Phi^\beta{}_{J\mathbf K_{p+1}}^{\mathbf K_{p}'}=
\nonumber\\
&&\qquad\qquad\quad
-s^\alpha_\beta{}_{J'\mathbf K_p'}^{\mathbf K_{p+1}}\nabla_{II'}
\big(\frac{p(p+5)}{2(p+2)}
[\overline\Phi^{J'K_p'},\Phi^{\mathbf K_{p-1}'}_{\mathbf K_{p+1}}]^\beta-
\nonumber\\
&&\qquad\qquad\qquad\qquad\qquad
\frac{3(1-\delta_{1p})}{p+2}
[\Phi_{K_{p}K_{p+1}},\overline{\Phi}_{\mathbf K_{p-1}}^{J'\mathbf K_{p-1}'}]\big)
+\Psi^\alpha_{2,II'}=
\label{E:SndDersTermSimp}\\
&&\qquad\qquad\quad
-s^\alpha_\beta{}_{J'\mathbf K_p'}^{\mathbf K_{p+1}}
\big(\frac{p(p+5)}{2(p+2)}
[\overline\Phi^{J'K_p'},\Phi^{\mathbf K_{p-1}'}_{II'\mathbf K_{p+1}}]^\beta-
\nonumber\\
&&\qquad\qquad\qquad\qquad\qquad
\frac{3(1-\delta_{1p})}{p+2}
[\Phi_{K_{p}K_{p+1}},\overline{\Phi}_{II'\mathbf K_{p-1}}^{J'\mathbf K_{p-1}'}]\big)
+\Psi^\alpha_{3,II'},\nonumber
\end{eqnarray}
where $\Psi^\alpha_{2,II'}$, $\Psi^\alpha_{3,II'}$ 
involve variables of order at most $p$. 
Moreover, by equation \eqref{E:wavePhiOrdp}, 
we have that
\begin{align}
s^\alpha_\beta{}_{I'\mathbf K_p'}^{\mathbf K_{p+1}}
\nabla^{JJ'}\nabla_{JJ'}\Phi^\beta{}_{I\mathbf K_{p+1}}^{\mathbf K_p'}&=
2(p+2)s^\alpha_\beta{}_{I'\mathbf K_p'}^{\mathbf K_{p+1}}
[\Phi_{S(I}^{\hp{K_p}},\Phi^{\mathbf K_p'S}_{\mathbf K_{p+1})}]^\beta+\nonumber\\
&2p\,s^\alpha_\beta{}_{I'\mathbf K_p'}^{\mathbf K_{p+1}}
[\overline\Phi^{K_p'}_{S'},\Phi^{\mathbf K_{p-1}'S'}_{\mathbf K_{p+1}I}]^\beta+
\Psi^\alpha_{4,II'},\label{E:WavesTerm}
\end{align}
where again $\Psi^\alpha_{4,II'}$ is of order $p$.
It follows from equations \eqref{E:sPhiordpPhi}, 
\eqref{E:SndDersTerm}, \eqref{E:SndDersTermSimp}, 
\eqref{E:WavesTerm} that
\begin{eqnarray}
&&\mbox{B}^o_{\rwseven,Z}\mbox{B}^p_{\rvseven,X,Y}\mathcal E^\alpha_{II'}(
s^\beta_\gamma{}_{J'\mathbf K_p'}^{\mathbf K_{p+1}}                     
\Phi^\gamma{}_{J\mathbf K_{p+1}}^{\mathbf K_{p}'})=
\nonumber\\
&&\qquad\qquad\qquad\qquad
-2(p+1)s^\alpha_\beta{}_{I'\mathbf K_p'}^{\mathbf K_{p+1}}
Z_{SK_{p+1}}X_{\mathbf K_p I}^S Y^{\mathbf K_p'} [\rw,\rv]^\beta.
\label{E:SndDersFinal}
\end{eqnarray}

Next we compute
\begin{eqnarray}
&&t^\alpha_\beta{}_{J\mathbf K_{p-1}'}^{\mathbf K_{p+2}}
\nabla^{JJ'}\nabla_{II'}\Phi^\beta{}_{J'\mathbf K_{p+2}}^{\mathbf K_{p-1}'}=
t^\alpha_\beta{}_{\mathbf K_{p-1}'}^{J\mathbf K_{p+2}}
\nabla_{II'}\nabla_{JJ'}\Phi^\beta{}_{\mathbf K_{p+2}}^{J'\mathbf K_{p-1}'}+
\nonumber\\
&&\qquad\qquad\qquad
t^\alpha_\beta{}_{\mathbf K_{p-1}'}^{J\mathbf K_{p+2}}
[\Phi_{IJ},\Phi_{I'\mathbf K_{p+2}}^{\mathbf K_{p-1}'}]^\beta+
t^\alpha_\beta{}_{I\mathbf K_{p-1}'}^{\mathbf K_{p+2}}
[\overline\Phi_{I'J'},\Phi_{\mathbf K_{p+2}}^{J'\mathbf K_{p-1}'}]^\beta.
\label{E:SndDertTerm}\end{eqnarray}
But by equation \eqref{E:derPhisymmvar},
\begin{eqnarray}
&&t^\alpha_\beta{}_{\mathbf K_{p-1}'}^{J\mathbf K_{p+2}}
\nabla_{II'}\nabla_{JJ'}\Phi^\beta{}_{\mathbf K_{p+2}}^{J'\mathbf K_{p-1}'}=
\nonumber\\
&&\qquad\qquad
\frac{p^2+p-2}{2p}\,t^\alpha_\beta{}_{\mathbf K_{p-1}'}^{J\mathbf K_{p+2}}
\nabla_{II'}[\Phi_{JK_{p+2}},\Phi_{\mathbf K_{p+1}}^{\mathbf K_{p-1}'}]^\beta
+\Psi^\alpha_{5,II'}=
\label{E:SndDertTermsSimpl}\\
&&\qquad\qquad\qquad\qquad
\frac{p^2+p-2}{2p}\,t^\alpha_\beta{}_{\mathbf K_{p-1}'}^{J\mathbf K_{p+2}}
[\Phi_{JK_{p+2}},\Phi_{II'\mathbf K_{p+1}}^{\mathbf K_{p-1}'}]^\beta
+\Psi^\alpha_{6,II'},
\nonumber\end{eqnarray}
where $\Psi^\alpha_{5,II'}$, $\Psi^\alpha_{6,II'}$ involve
variables of order at most $p$. Moreover, by 
equation \eqref{E:wavePhiOrdp}, we have that
\begin{align}
t^\alpha_\beta{}_{I\mathbf K_{p-1}'}^{\mathbf K_{p+2}}
\nabla^{JJ'}\nabla_{JJ'}\Phi^\beta{}_{I'\mathbf K_{p+2}}^{\mathbf K_{p-1}'}&=
t^\alpha_\beta{}_{I\mathbf K_{p-1}'}^{\mathbf K_{p+2}}\big(
2(p+2)[\Phi_{SK_{p+2}},\Phi^{\mathbf K_{p-1}'S}_{I'\mathbf K_{p+1}}]^\beta+
\nonumber\\
&\qquad 2p[\overline\Phi_{S'}^{(K_p'},
\Phi_{\mathbf K_{p+2}}^{\mathbf K_{p-1}')S'}]^\beta
\e_{K_p'I'}\big)+\Psi^\alpha_{7,II'},\label{E:WavetTerm}
\end{align}
where again $\Psi^\alpha_{7,II'}$ is of order $p$. 
It follows from equations
\eqref{E:tPhiordpPhi}, \eqref{E:SndDertTerm}, 
\eqref{E:SndDertTermsSimpl}, \eqref{E:WavetTerm} that
\begin{eqnarray}
&&\mbox{B}^o_{\rwseven,Z}\mbox{B}^p_{\rvseven,X,Y}\mathcal E^\alpha_{II'}(
t^\beta_\gamma{}_{I\mathbf K_{p-1}'}^{\mathbf K_{p+2}}
\Phi^\gamma{}_{I'\mathbf K_{p+2}}^{\mathbf K_{p-1}'}) =
\nonumber\\
&&\qquad\quad\big(\frac{p^2+p-2}{2p}\,
t^\alpha_\beta{}^{\mathbf K_{p+3}}_{\mathbf K_{p-1}'}
Z_{K_{p+3}K_{p+2}}X_{I}+
2t^\alpha_\beta{}^{\mathbf K_{p+3}}_{\mathbf K_{p-1}'}
Z_{IK_{p+3}}X_{K_{p+2}}-\label{E:SndDertFinal}\\
&&\qquad\qquad\qquad
2(p+2)t^\alpha_\beta{}^{\mathbf K_{p+2}}_{I\mathbf K_{p-1}'}
Z_{SK_{p+2}}X^S\big)
X_{\mathbf K_{p+1}}Y_{I'}^{\mathbf K_{p-1}'}[\rw,\rv]^\beta.
\nonumber
\end{eqnarray}

One can easily verify that
\begin{align}
\mbox{B}^o_{\rwseven,Z}\mbox{B}^p_{\rvseven,X,Y}\mathcal E^\alpha_{II'}(
\overline s^\beta_\gamma{}_{J\mathbf K_p}^{\mathbf K_{p+1}'}                     
\Phibar^\gamma{}_{J'\mathbf K_{p+1}'}^{\mathbf K_{p}})=0,
\label{E:ConjTerms1}\\
\mbox{B}^o_{\rwseven,Z}\mbox{B}^p_{\rvseven,X,Y}\mathcal E^\alpha_{II'}(
\overline t^\beta_\gamma{}_{I'\mathbf K_{p-1}}^{\mathbf K_{p+2}'}
\Phibar^\gamma{}_{I\mathbf K_{p+2}'}^{\mathbf K_{p-1}})=0.
\label{E:ConjTerms2}
\end{align}

Finally, by equations \eqref{E:zPhiordpPhi}, \eqref{E:qPhiordpPhi}, 
\eqref{E:PhiOrdp1Phi}, \eqref{E:SndDersFinal}, \eqref{E:SndDertFinal},
\eqref{E:ConjTerms1}, \eqref{E:ConjTerms2}, we see that
\begin{eqnarray}
&&X^I\mbox{B}^o_{\rwseven,Z}\mbox{B}^p_{\rvseven,X,Y}
\mathcal E^\alpha_{II'}(Q^\beta_{JJ'}) =\nonumber\\ 
&&\qquad\qquad
2(p+3)t^\alpha_\beta{}^{\mathbf K_{p+3}}_{\mathbf K_{p-1}'}
Z_{K_{p+3}}X_{\mathbf K_{p+2}}Y_{I'}^{\mathbf K_{p-1}'}
Z_S X^S[\rw,\rv]^\beta=0,\nonumber\\
&&Y^{I'}\mbox{B}^o_{\rwseven,Z}\mbox{B}^p_{\rvseven,X,Y}
\mathcal E^\alpha_{II'}(Q^\beta_{JJ'}) =\nonumber\\
&&\qquad\qquad
-2(p+1)s^\alpha_\beta{}_{\mathbf K_{p+1}'}^{\mathbf K_{p+1}}
Z_{K_{p+1}}X_{\mathbf K_p I} Y^{\mathbf K_{p+1}'}Z_S X^S [\rw,\rv]^\beta=0,
\nonumber
\end{eqnarray}
from which it immediately follows that
\[
s^\alpha_\beta{}_{\mathbf K_{p+1}'}^{\mathbf K_{p+1}}=0,\qquad
t^\alpha_\beta{}^{\mathbf K_{p+3}}_{\mathbf K_{p-1}'}=0,
\]
as required.\end{proof}

\begin{proposition}\label{P:gDerivations}
Let $\fg$ be a 
semi-simple Lie algebra with structure 
constants $c^\alpha_{\beta\gamma}$ 
in a basis $\{e_\alpha\}$.
Suppose that 
$\rz^\alpha_\beta=\rz^\alpha_\beta(x^j)$ 
are smooth functions satisfying 
\[\rz^\alpha_\beta c^\beta_{\gamma\delta}+
\rz^\beta_\delta c^\alpha_{\beta\gamma}-
\rz^\beta_{\gamma}c^\alpha_{\beta\delta} = 0.\]
Then there is a $\fg$-valued smooth function 
$\rw^\alpha=\rw^\alpha(x^j)$ such that 
\begin{equation}
\rz^\alpha_\beta = c^\alpha_{\beta\gamma}
\rw^\gamma.
\label{E:gDerivations}\end{equation} 
\end{proposition}
\begin{proof}
Let $Z=Z(x^j)$ be the $\mbox{End}(\fg)$-valued 
function defined by 
\[Z(x^j)e_\alpha =\rz^\beta_\alpha(x^j) e_\beta.\]
Then the assumptions imply that
\[Z(x^j)[\rv_1,\rv_2] = [Z(x^j)\rv_1,\rv_2]+[\rv_1,Z(x^j)\rv_2],\]
that is, $Z(x^j)$ is a derivation of $\fg$
for every $x^j$.
Since $\fg$ is semi-simple, any derivation is
inner, that is, there is a $\fg$-valued function
$\rw^\alpha=\rw^\alpha(x^j)$ such that
\[Z(x^j)\rv = [\rv,\rw(x^j)]\qquad\mbox{for all $\rv\in \fg$}.\] 
This implies that 
\[
\rw^\alpha(x^j) = -\kappa^{\alpha\beta}
c_{\beta\gamma}^\delta \rz^\gamma_\delta(x^j),
\]
where $\kappa^{\alpha\beta}$ is the inverse
of the Killing form 
$\kappa_{\alpha\beta} = c_{\alpha\gamma}^\delta 
c_{\beta\delta}^\gamma$. Thus $\rw^\alpha(x^j)$
is smooth.\end{proof}

\begin{bew}{of Theorem \ref{T:maintheorem}}
By applying Lemmas \ref{L:aReduction}, 
\dots, \ref{L:PhiordpPhi}
repeatedly, we see that any generalized symmetry 
of the Yang-Mills equations
is equivalent to a first order symmetry 
$\wQ$ with components $\wQ^\alpha_{II'}$
of the form
\begin{equation}
\label{E:FirstOrderQ}
\wQ^\alpha_{II'}=q^\alpha_\beta(x^j)a^\beta_{II'}+
s^\alpha_\beta{}^{K}_{I'}(x^j)\Phi^\beta_{IK}+
\overline s^\alpha_\beta{}^{K'}_{I}(x^j)
\overline\Phi^\beta_{I'K'}+
v^\alpha_{II'}(x^j).
\end{equation}
On $\mathcal R$ we have that
\[
D_i(q^\alpha_\beta a^\beta_{j})-
D_j(q^\alpha_\beta a^\beta_{i})=
(\partial_i q^\alpha_\beta) a^\beta_j-
(\partial_j q^\alpha_\beta) a^\beta_i+
q^\alpha_\beta(F^\beta_{ij} -
c^\beta_{\gamma\delta}a^\gamma_i a^\delta_j),
\]
and
\begin{eqnarray}
&&D^j(D_i(q^\alpha_\beta a^\beta_{j})-
D_j(q^\alpha_\beta a^\beta_{i}))=
(\partial^j\partial_i q^\alpha_\beta) a^\beta_j-
(\partial^j\partial_j q^\alpha_\beta) a^\beta_i+
\nonumber\\
&&\qquad\qquad(\partial_i q^\alpha_\beta) a^\beta{}^j_j -
\partial^j q^\alpha_\beta(a^\beta_{ij}-
\frac{3}{2}F^\beta_{ij} +
\frac{3}{2}c^\beta_{\gamma\delta}a^\gamma_i a^\delta_j)-
\label{E:SndDerqa}\\
&&\qquad\qquad\qquad 
q^\alpha_\beta c^\beta_{\gamma\delta}(a^\gamma{}^j F^\delta_{ij}+
(a^\gamma_{ij}-\frac{1}{2}F^\gamma_{ij}+
\frac{1}{2}c^\gamma_{\zeta\eta}a^\zeta_i a^\eta_j) a^\delta{}^j
+ a^\gamma_i a^\delta{}_j^j).\nonumber
\end{eqnarray}

Consequently, the only terms involving the variables
$\partial^1\Phi$
in the determining equations for $\wQ$ arise
from the term $s^\alpha_\beta{}^{K}_{I'}(x^j)\Phi^\beta_{IK}$
and its complex conjugate. Hence we can repeat the
computations leading to equations \eqref{E:Killingspinor}, 
\eqref{E:stadjcomm2} in the proof of Lemma \ref{L:ordpPhired} 
to see that the $s^\alpha_\beta{}^K_{K'}$ are Killing spinors 
satisfying
\begin{equation}
\label{E:Ord1sAdjComm}
s^\alpha_\beta{}^K_{K'} c^\beta_{\gamma\delta}+
c^\alpha_{\beta\gamma} s^\beta_\delta{}^K_{K'}=0.
\end{equation}

Our next goal is to show that $s^\alpha_\beta{}^K_{K'}$ 
must be real. For this, we analyze terms quadratic
in the variables $\Phi$, $\overline\Phi$ in the
determining equations for $\wQ$.
These only arise from the second order covariant 
derivatives of $\Phi$, $\overline\Phi$ and from 
the bracket term in \eqref{E:YMsymmspinor}. 
Note that
\[
\nabla_{II'}\Phi_{JK} = \Phi_{II'JK}\qquad\mbox{on $\mathcal R$}.
\]
We use the above equation and \eqref{E:derPhisymmvar} to compute
\begin{align}
s^\alpha_\beta{}_{J'}^K\nabla^{JJ'}\nabla_{II'}\Phi^\beta_{JK}&=
-s^\alpha_\beta{}_{I'}^K[\Phi_I^L,\Phi_{KL}]^\beta+
s^\alpha_\beta{}_{J'}^K[\Phi_{IK},\overline\Phi_{I'}^{J'}]^\beta,
\label{E:SndDersTerms}\\
s^\alpha_\beta{}_{I'}^K\nabla^{JJ'}\nabla_{JJ'}\Phi^\beta_{IK}&=
-2s^\alpha_\beta{}_{I'}^K[\Phi_I^L,\Phi_{KL}]^\beta\label{E:WavesTerms}
\end{align}
on $\mathcal R^1$. 
Moreover, on account of \eqref{E:Ord1sAdjComm}, we have that
\begin{equation}
c^\alpha_{\beta\gamma}F^\beta_{II'}{}^{JJ'}
s^\gamma_\delta{}^{K}_{J'}\Phi^\delta_{JK}=
s^\alpha_\beta{}_{I'}^K[\Phi_I^J,\Phi_{JK}]^\beta +
s^\alpha_\beta{}_{J'}^K[\overline\Phi_{I'}^{J'},\Phi_{IK}]^\beta.
\label{E:sBracketTerms}
\end{equation}

It follows from \eqref{E:SndDersTerms}, \eqref{E:WavesTerms},
\eqref{E:sBracketTerms} and their complex 
conjugate equations  that the determining equations 
for $\wQ$ yield the terms 
\[
2s^\alpha_\beta{}_{J'}^K[\Phi_{IK},\overline\Phi_{I'}^{J'}]^\beta-
2\overline s^\alpha_\beta{}_J^{K'}[\Phi_I^J,\overline\Phi_{I'K'}]^\beta
\]
quadratic in $\Phi$, $\overline\Phi$. 
These must vanish and hence
\[
s^\alpha_\beta{}^{K}_{K'} = 
\overline s^\alpha_\beta{}^{K}_{K'},
\] 
that is, $s^\alpha_\beta{}^{K}_{K'}$ is real.

Next by \eqref{E:Ord1sAdjComm}, for fixed $K$, $K'$, the 
endomorphisms $S_{K'}^K$ of $\fg$ defined by 
$S_{K'}^K(e_\alpha) = s_\alpha^\beta{}_{K'}^Ke_\beta$
commute with the adjoint representation of $\fg$.
Write $\fg=\fg_1+\cdots + \fg_n$, where $\fg_m\subset \fg$
is a simple ideal. By Proposition \ref{P:adjointcommute}, 
$S_{K'}^K$  leaves each ideal $\fg_m$ invariant
and the restriction $S_{m,}{}_{K'}^{K}$ of $S_{K'}^K$
to $\fg_m$ is either a multiple of the identity mapping
or a linear combination of the identity mapping and
the almost complex structure $J_{m}$
of $\fg_m$. It follows that the terms 
$s^\alpha_\beta{}^{K}_{I'}\Phi^\beta_{IK}+
\overline s^\alpha_\beta{}^{K'}_{I}
\overline\Phi^\beta_{I'K'}$ in
$\wQ^\alpha_{II'}$ agree with
the components of a sum
of the symmetries $Q_{m}[\xi]$,
${Q}_{J,m}[\tau]$ defined in \eqref{E:confJsymm}.
Thus we can subtract this sum
from $\wQ$ to get a symmetry $Q$ with
components
\begin{equation}
\label{E:Qw/osTerms}
Q^\alpha_{i} =  q^\alpha_\beta(x^j)a^\beta_{i}+
v^\alpha_i(x^j).
\end{equation}

Our next goal is to show that $Q$ 
must be a gauge symmetry. For this
collect terms in the determining equations for 
$Q$ involving the product $a\,\partial^1 a$.
With the help of \eqref{E:SndDerqa} we get the equation
\begin{align}
-q^\alpha_\beta c^\beta_{\gamma\delta} 
a^\gamma{}_i^j a^\delta_j-
&q^\alpha_\beta c^\beta_{\gamma\delta} 
a^\gamma_i a^\delta{}_j^j-
c^\alpha_{\beta\gamma} a^\beta{}^j 
q^\gamma_\delta a^\delta_{ij}+
\nonumber\\
&c^\alpha_{\beta\gamma} a^\beta_i 
q^\gamma_\delta a^\delta{}_j^j -
c^\alpha_{\beta\gamma} a^\beta{}_j^j 
q^\gamma_\delta a^\delta_i+
c^\alpha_{\beta\gamma} a^\beta{}_i^j 
q^\gamma_\delta a^\delta_j = 0,
\nonumber
\end{align}
which simplifies to
\[
(q^\alpha_\beta c^\beta_{\gamma\delta}+
 q^\beta_\delta c^\alpha_{\beta\gamma}-
 q^\beta_\gamma c^\alpha_{\beta\delta})
(a_i^\gamma a^\delta{}_j^j-a^\gamma_j a^\delta{}_i^j)=0.
\]
Hence we have 
\[
q^\alpha_\beta c^\beta_{\gamma\delta}+
 q^\beta_\delta c^\alpha_{\beta\gamma}-
 q^\beta_\gamma c^\alpha_{\beta\delta}=0,
\]
which, by Proposition \ref{P:gDerivations}, implies that
\begin{equation}
\label{E:qAsDerivation}
q^\alpha_\beta(x^j) = 
c^\alpha_{\beta\gamma}w^\gamma(x^j),
\end{equation}
for some $w^\gamma = w^\gamma(x^j)$.
Next collect terms involving $\partial^1 a$ 
in the determining equations for $Q$.
This yields the equation
\[
c^\alpha_{\beta\gamma}
(\partial_i w^\gamma - v^\gamma_i)a^\beta{}_j^j -
c^\alpha_{\beta\gamma}
(\partial_j w^\gamma - v^\gamma_j)a^\beta{}_i^j=0,
\]
from which it follows that
\begin{equation}
\label{E:vAsGradient}
v^\gamma_i = \partial_i w^\gamma.
\end{equation}
By \eqref{E:Qw/osTerms}, 
\eqref{E:qAsDerivation}, \eqref{E:vAsGradient}, 
$Q$ is a gauge symmetry, 
as required.
\end{bew}

\appendix
\section{}
\label{A:appendix}
In this Section we prove the second part 
of Proposition \ref{P:dersymmvar}. We start 
by collecting together a few identities 
needed in the course of the proof.

The commutation formula \eqref{E:covarcommutespin}
implies that
\begin{equation}
\label{E:SymmCovarDer}
\nabla_{(I}^{[I'}\nabla_{J)}^{J']}G
= \frac{1}{2}\e^{I'J'}[\Phi_{IJ},G],\quad
\nabla_{[I}^{(I'}\nabla_{J]}^{J')}G
= \frac{1}{2}\e_{IJ}[\overline \Phi^{I'J'},G],
\end{equation}
and
\begin{equation}
\label{E:SkewCovarDer}
\nabla_{K'(I}\nabla_{J)}^{K'}G = [\Phi_{IJ},G],\qquad
\nabla_{K(I'}\nabla_{J')}^{K}G = [\overline\Phi_{I'J'},G].
\end{equation}
Write 
\[\nabla^2 = \nabla_{AA'}\nabla^{AA'},\qquad
\Delta_{II'}=\nabla_{I'}^J\Phi_{IJ}\] 
for the wave operator and the 
Yang-Mills equations in spinor form. 
\begin{proposition}\label{P:WaveCommute}
Let $G$ be a 
$\fg$-valued differential function and
let $\nabla^2$ be the wave operator. 
Then, on solutions of the Yang-Mills
equations,
\begin{eqnarray}
&&\mbox{\rm (i)}\qquad\nabla^2\Phi_{IJ} = 
2[\Phi_{IK},\Phi_J^K],\label{E:WavePhi}\\
&&\mbox{\rm (ii)}\qquad(\nabla^2\nabla_{II'}-
\nabla_{II'}\nabla^2)G=
2[\Phi_{IJ},\nabla_{I'}^{J}G]+
2[\overline\Phi_{I'J'},\nabla_{I}^{J'}G].\label{E:WaveCommuteFinal}
\end{eqnarray}
\end{proposition}
\begin{proof} We use \eqref{E:SkewCovarDer}
to compute
\begin{eqnarray}
&&\nabla_{K'I}\Delta_{J}^{K'} = 
\nabla_{K'I}\nabla^{KK'}\Phi_{JK} =\nonumber\\
&&-\nabla_{K'(I}\nabla_{K)}^{K'}\Phi_{J}^K-
\frac{1}{2}\e_{IK}\nabla^2\Phi_{J}^K=
-[\Phi_{IK},\Phi_J^K]+\frac{1}{2}\nabla^2\Phi_{IJ},
\nonumber\end{eqnarray}
from which (i) follows. 

Next we use \eqref{E:covarcommutespin} to compute
\[
\nabla^{JJ'}\nabla_{II'}G = \nabla_{II'}\nabla^{JJ'}G +
\e^{J'}{}_{I'}[\Phi_I^J,G]+\e^J{}_I[\overline\Phi_{I'}^{J'},G].
\]
Hence
\begin{eqnarray}
&&\nabla_{JJ'}\nabla^{JJ'}\nabla_{II'}G =
\nabla_{JJ'}\nabla_{II'}\nabla^{JJ'}G -
\nabla_{JI'}[\Phi_I^J,G] - \nabla_{IJ'}[\overline\Phi_{I'}^{J'},G]=
\nonumber\\
&&\nabla_{JJ'}\nabla_{II'}\nabla^{JJ'}G-
[\Phi_I^J,\nabla_{JI'} G] - [\overline\Phi_{I'}^{J'},\nabla_{IJ'} G] +
[\Delta_{II'},G] +[\overline \Delta_{II'},G]. 
\label{E:WaveCommute}\end{eqnarray}
But by \eqref{E:covarcommutespin},
\begin{equation}
\nabla_{JJ'}\nabla_{II'}\nabla^{JJ'}G=
\nabla_{II'}\nabla_{JJ'}\nabla^{JJ'}G
+[\Phi_{IJ},\nabla_{I'}^{J}G]+
[\overline\Phi_{I'J'},\nabla_{I}^{J'}G].
\label{E:AnotherCommFrml}\end{equation}
Now (ii) follows from equations
\eqref{E:WaveCommute}, \eqref{E:AnotherCommFrml}.
\end{proof}

\begin{bew}{of Proposition \ref{P:dersymmvar} \rm{(ii)}}
We first prove \eqref{E:wavePhiOrdp}.
By \eqref{E:WaveCommuteFinal},
\begin{align}
\nabla^2\Pvar{K}{}{p+2}{p}=
&\nabla^2\nabla^{(K_{1}'}_{(K_{1}}\cdots
\nabla^{K_p')}_{K_p}\Phi_{K_{p+1}K_{p+2})}=\nonumber\\
&\nabla^{(K_{1}'}_{(K_{1}}\nabla^2
\nabla^{K_{2}'}_{K_{2}}\cdots
\nabla^{K_p')}_{K_p}\Phi_{K_{p+1}K_{p+2})} +
2[\Phi_{S(K_{p+2}},\Phi^{\mathbf K_{p}'S}_{\mathbf K_{p+1})}]+
\nonumber\\
&\qquad\qquad\qquad\qquad\qquad
2[\overline\Phi_{S'}^{(K_{p}'},
\Phi_{\mathbf K_{p+2}}^{\mathbf K_{p-1}')S'}]+
\widehat{\mathcal W}_{1}= \cdots = \nonumber\\
&\nabla^{(K_{1}'}_{(K_{1}}
\cdots\nabla^{K_p')}_{K_p}
\nabla^2\Phi_{K_{p+1}K_{p+2})}+
2p[\Phi_{S(K_{p+2}},
\Phi^{\mathbf K_{p}'S}_{\mathbf K_{p+1})}]+
\label{E:WaveOrdp-1Phi}\\
&\qquad\qquad\qquad\qquad\qquad
2p[\overline\Phi_{S'}^{(K_{p}'},
\Phi_{\mathbf K_{p+2}}^{\mathbf K_{p-1}')S'}]+
\widehat{\mathcal W}_{2}=\nonumber\\
&2(p+2-\delta_{0p})[\Phi_{S(K_{p+2}},
\Phi^{\mathbf K_{p}'S}_{\mathbf K_{p+1})}]+
2p[\overline\Phi_{S'}^{(K_{p}'},
\Phi_{\mathbf K_{p+2}}^{\mathbf K_{p-1}')S'}]+
\widehat{\mathcal W}_3,\nonumber
\end{align}
where $\widehat{\mathcal W}_{1}$,
$\widehat{\mathcal W}_{2}$, $\widehat{\mathcal W}_3$,
when restricted to $\mathcal R^{p}$,
depend on the variables $\Phi^{[p-1]}$. This proves
\eqref{E:wavePhiOrdp}.

In order to prove \eqref{E:derPhisymmvar} 
we first reduce the derivative 
$\nabla_{K_{p+3}}^{K_{p+1}'}\Pvar{K}{}{p+2}{p}$
into symmetric components to derive the expression
\begin{eqnarray}
&&\nabla_{K_{p+3}}^{K_{p+1}'}\Pvar{K}{}{p+2}{p} =
\Pvar{K}{}{p+3}{p+1} +
\frac{p}{p+1}\e^{K_{p+1}'(K_p'}
\nabla_{S'(K_{p+3}}\Phi^{\mathbf K_{p-1}')S'}_{\mathbf K_{p+2})}-
\nonumber\\
&&\qquad\qquad\frac{p+2}{p+3}\e_{K_{p+3}(K_{p+2}}
\nabla^{S(K_{p+1}'}\Phi^{\mathbf K_p')}_{\mathbf K_{p+1})S} -
\label{E:RedSymmComp}\\
&&\qquad\qquad\qquad\qquad\frac{p(p+2)}{(p+1)(p+3)}
\e^{K_{p+1}'(K_p'}\e_{K_{p+3}(K_{p+2}}
\nabla^{|S|}_{|S'|}\Phi^{\mathbf K_{p-1}')S'}_{\mathbf K_{p+1})S}.
\nonumber\end{eqnarray}
We simplify the terms 
$\nabla_{S'(K_{p+3}}\Phi^{\mathbf K_{p-1}'S'}_{\mathbf K_{p+2})}$,
$\nabla^{S(K_{p+1}'}\Phi^{\mathbf K_p')}_{\mathbf K_{p+1}S}$,
$\nabla^{S}_{S'}\Phi^{\mathbf K_{p-1}'S'}_{\mathbf K_{p+1}S}$
in \eqref{E:RedSymmComp} separately. 

First, we have that
\begin{eqnarray}
&&\nabla_{S'(K_{p+3}}\Phi^{\mathbf K_{p-1}'S'}_{\mathbf K_{p+2})}=
\nabla_{S'(K_{p+3}}\nabla^{(K_{1}'}_{K_{1}}\cdots
\nabla^{K_{p-1}'}_{K_{p-1}}\nabla^{S')}_{K_p}\Phi_{K_{p+1}K_{p+2})}=\nonumber\\
&&\qquad\qquad\frac{1}{p}\sum_{s=1}^p
\nabla_{S'(K_{p+3}}\nabla_{K_1}^{(K_{1}'}\cdots
\nabla^{K_{s-1}'}_{K_{s-1}}
\nabla^{|S'|}_{K_s}\nabla^{K_{s}'}_{K_{s+1}}\cdots
\nabla^{K_{p-1}')}_{K_p}\Phi_{K_{p+1}K_{p+2})}.
\nonumber\end{eqnarray}
Write
\[
\mathcal A_s = 
\nabla_{S'(K_{p+3}}\nabla_{K_1}^{(K_{1}'}\cdots
\nabla^{K_{s-1}'}_{K_{s-1}}
\nabla^{|S'|}_{K_s}\nabla^{K_{s}'}_{K_{s+1}}\cdots
\nabla^{K_{p-1}')}_{K_p}\Phi_{K_{p+1}K_{p+2})}
\]
so that
\begin{eqnarray}
\label{E:Asum}
\nabla_{S'(K_{p+3}}
\Phi^{\mathbf K_{p-1}'S'}_{\mathbf K_{p+2})}=
\frac{1}{p}\sum_{s=1}^p \mathcal A_s.
\end{eqnarray}
Then, by \eqref{E:SymmCovarDer},
\begin{eqnarray}
&&\mathcal A_s = \nabla_{S'(K_{p+3}}\nabla_{K_1}^{(K_{1}'}
\cdots\nabla^{K_{s-2}'}_{K_{s-2}}
\nabla^{|S'|}_{K_{s-1}}\nabla^{K_{s-1}'}_{K_{s}}
\cdots\nabla^{K_{p-1}')}_{K_p}
\Phi_{K_{p+1}K_{p+2})}+\nonumber\\
&&\nabla_{S'(K_{p+3}}\nabla_{K_1}^{(K_{1}'}
\cdots\nabla^{K_{s-2}'}_{K_{s-2}}
(\e^{K_{s-1}'|S'|}[\Phi_{K_{s-1}K_s},\nabla^{K_{s}'}_{K_{s+1}}
\cdots\nabla^{K_{p-1}')}_{K_p}
\Phi_{K_{p+1}K_{p+2})}])=
\label{E:ATermsCommute}\\
&&\mathcal A_{s-1}+\nabla_{(K_{p+3}}^{(K_{1}'}
\nabla_{K_{1}}^{K_{2}'}
\cdots\nabla^{K_{s-1}'}_{K_{s-2}}
[\Phi_{K_{s-1}K_s},\nabla^{K_{s}'}_{K_{s+1}}
\cdots\nabla^{K_{p-1}')}_{K_p}
\Phi_{K_{p+1}K_{p+2})}] =\nonumber\\
&&\mathcal A_{s-1}+(1-\delta_{sp})[\Phi_{(K_{p+3}K_{p+2}},
\Phi^{\mathbf K_{p-1}'}_{\mathbf K_{p+1})}]+
\widehat{\mathcal A}_{o,s},
\nonumber
\end{eqnarray}
where $\widehat{\mathcal A}_{o,s}$ 
is a function of the variables $\Phi^{[p-2]}$. 
Now a repeated application of 
\eqref{E:ATermsCommute} yields the equation
\begin{equation}
\label{E:AsInTermsAp}
\mathcal A_s = \mathcal A_1 + 
(s-1-\delta_{sp})[\Phi_{(K_{p+3}K_{p+2}},
\Phi^{\mathbf K_{p-1}'}_{\mathbf K_{p+1})}] +
\widehat{\mathcal A}_s,
\end{equation}
where $\widehat{\mathcal A}_{s}$ 
is a function of the variables $\Phi^{[p-2]}$.
By equation \eqref{E:SkewCovarDer},
\begin{equation}
\label{E:ApTerm}
\mathcal A_1 = [\Phi_{(K_{p+3}K_{p+2}},
\Phi^{\mathbf K_{p-1}'}_{\mathbf K_{p+1})}].
\end{equation}
Consequently, equations \eqref{E:Asum}, 
\eqref{E:AsInTermsAp}, 
\eqref{E:ApTerm} imply that
\begin{equation}
\nabla_{S'(K_{p+3}}
\Phi^{\mathbf K_{p-1}')S'}_{\mathbf K_{p+2})}=
\frac{(p-1)(p+2)}{2p}[\Phi_{(K_{p+3}K_{p+2}},
\Phi^{\mathbf K_{p-1}'}_{\mathbf K_{p+1})}]+
\widehat{\mathcal A},\label{E:ATermFinal}
\end{equation}
where $\widehat{\mathcal A}$ is a function of 
the variables $\Phi^{[p-2]}$.

Next note that
\begin{eqnarray}
&&\nabla^{S(K_{p+1}'}\Phi^{\mathbf K_p')}_{\mathbf K_{p+1}S}=
\frac{2}{p+2}\nabla^{S(K_{p+1}'}\nabla^{K_1'}_{(K_{1}}\cdots
\nabla^{K_{p}')}_{K_{p}}\Phi_{K_{p+1})S}+\nonumber\\
&&\quad\frac{1}{p+2}\sum_{s=1}^p
\nabla^{S(K_{p+1}'}\nabla^{K_1'}_{(K_{1}}
\cdots\nabla^{K_{s-1}'}_{K_{s-1}}
\nabla_{|S|}^{K_s'}\nabla^{K_{s+1}'}_{K_{s}}\cdots
\nabla^{K_p')}_{K_{p-1}}\Phi_{K_pK_{p+1})}.
\label{E:PrimeSymmTerm}
\end{eqnarray}
Write
\begin{equation}
\label{E:BsDefn}
\mathcal B_s =
\nabla^{S(K_{p+1}'}\nabla^{K_1'}_{(K_{1}}
\cdots\nabla^{K_{s-1}'}_{K_{s-1}}
\nabla_{|S|}^{K_s'}\nabla^{K_{s+1}'}_{K_{s}}\cdots
\nabla^{K_p')}_{K_{p-1}}\Phi_{K_pK_{p+1})}.
\end{equation}
Note that
\begin{equation}
\mathcal B_1 = -[\overline\Phi^{K_1'K_2'},\Phi_{K_1K2}],
\qquad\mbox{if $p=1$}.\label{E:B1Term}
\end{equation}
When $p\geq2$, we use \eqref{E:SymmCovarDer} to compute
\begin{eqnarray}
&&\mathcal B_s = \mathcal B_{s-1}-\nabla_{(K_{1}}^{(K_{1}'}\cdots
\nabla^{K_{s-1}'}_{K_{s-1}}[\overline\Phi^{K_{s}'K_{s+1}'},
\nabla^{K_{s+2}'}_{K_{s}}
\cdots\nabla^{K_{p+1'})}_{K_{p-1}}\Phi_{K_pK_{p+1})}]=\nonumber\\
&&\qquad\qquad
\mathcal B_{s-1} - [\overline\Phi^{(K_{p+1}'K_p'},
\Phi_{\mathbf K_{p+1}}^{\mathbf K_{p-1}')}]- 
\delta_{sp}[\overline\Phi^{\mathbf K_{p+1}'}_{(\mathbf K_{p-1}},
\Phi_{K_pK_{p+1})}]+\widehat {\mathcal B}_{o,s},\nonumber
\end{eqnarray}
where $\widehat {\mathcal B}_{o,s}$ is a function 
of the variables $\Phi^{[p-2]}$. Hence
\begin{equation}
\label{E:BsInTermsBp}
\mathcal B_s = \mathcal B_1 - 
(s-1)[\overline\Phi^{(K_{p+1}'K_p'},
\Phi_{\mathbf K_{p+1}}^{\mathbf K_{p-1}')}]-
\delta_{sp}[\overline\Phi^{\mathbf K_{p+1}'}_{(\mathbf K_{p-1}},
\Phi_{K_pK_{p+1})}]+\widehat {\mathcal B}_{s},
\end{equation}
where $\widehat {\mathcal B}_{s}$ is a 
function of the variables $\Phi^{[p-2]}$.

By virtue of the identity
\[
\nabla^{K_{p}'}_{K_p}\Phi_{K_{p+1}S} = 
\nabla^{K_{p}'}_{S}\Phi_{K_{p+1}K_p} +
\e_{SK_p}\Delta_{K_{p+1}}^{K_p'}
\]
we have that
\begin{equation}
\label{E:SonPhiTerm}
\nabla^{S(K_{p+1}'}\nabla^{K_1'}_{(K_{1}}\cdots
\nabla^{K_{p}')}_{K_p}\Phi_{K_{p+1})S}= 
\mathcal B_p + \nabla^{(K_{1}'}_{(K_{1}}\cdots
\nabla^{K_{p}'}_{K_p}\Delta^{K_{p+1}')}_{K_{p+1})}.
\end{equation}
Clearly
\begin{equation}
\label{E:BpTerm}
\mathcal B_1 = - [\overline\Phi^{(K_{p+1}'K_p'},
\Phi^{\mathbf K_{p-1}')}_{\mathbf K_{p+1}}].
\end{equation}
Consequently, equations 
\eqref{E:PrimeSymmTerm},\dots,\eqref{E:BpTerm}
together imply that
\begin{eqnarray}
&&\nabla^{S(K_{p+1}'}\Phi^{\mathbf K_p')}_{\mathbf K_{p+1})S}=
-\frac{p(p+5)}{2(p+2)}
[\overline{\Phi}^{(K_{p+1}'K_p'},\Phi^{\mathbf K_{p-1}')}_{\mathbf K_{p+1}}]-
\nonumber\\
&&\qquad\qquad\qquad\qquad\qquad
\frac{3(1-\delta_{1p})}{p+2}
[\overline{\Phi}^{\mathbf K_{p+1}'}_{(\mathbf K_{p-1}},
\Phi_{K_pK_{p+1})}]+\widehat{\mathcal B},\label{E:BTermsFinal}
\end{eqnarray}
where $\widehat{\mathcal B}$, when 
restricted to $\mathcal R^{p-1}$, is a 
function of the variables $\Phi^{[p-2]}$.

Next note that
\[
\nabla^S_{S'}\Phi^{\mathbf K_{p-1}'S'}_{\mathbf K_{p+1}S}=
\frac{1}{p}\sum_{s=1}^p 
\nabla^S_{S'} \nabla^{(K_{1}'}_{(K_1}\cdots 
\nabla^{K_{s-1}'}_{K_{s-1}}
\nabla^{|S'|}_{K_s}\nabla^{K_s'}_{K_{s+1}}\cdots
\nabla^{K_{p-1}')}_{K_p}\Phi_{K_{p+1}S)}.
\]
Write
\[
\mathcal C_s = \nabla^S_{S'} \nabla^{(K_{1}'}_{(K_1}
\cdots \nabla^{K_{s-1}'}_{K_{s-1}}\nabla^{|S'|}_{K_s}
\nabla^{K_s'}_{K_{s+1}}\cdots
\nabla^{K_{p-1}')}_{K_p}\Phi_{K_{p+1}S)}.
\]
Suppose that $p\geq2$. Then
\begin{align}
\mathcal C_s = &\mathcal C_{s-1}+\nabla^{S(K_{1}'}
\nabla^{K_{2}'}_{(K_1}\cdots
\nabla^{K_{s-1}'}_{K_{s-2}}[\Phi_{K_{s-1}K_s},
\nabla^{K_s'}_{K_{s+1}}\cdots \nabla_{K_p}^{K_{p-1}'}
\Phi_{K_{p+1}S)}]=
\nonumber\\
&\mathcal C_{s-1}+(1-\delta_{sp})[\Phi_{(K_{p+1}K_p},
\Phi^{\mathbf K_{p-1}'S}_{\mathbf K_{p-1}S)}]+
\widehat{\mathcal C}_{s,o}=
\nonumber\\
&\mathcal C_{s-1}+\frac{2}{p+2}(1-\delta_{sp})[\Phi_{S(K_{p+1}},
\Phi^{\mathbf K_{p-1}'S}_{\mathbf K_{p})}]+
\widehat{\mathcal C}_{s,o} = \cdots = \nonumber\\
&\mathcal C_{1}+\frac{2}{p+2}(s-1-\delta_{sp})[\Phi_{S(K_{p+1}},
\Phi^{\mathbf K_{p-1}'S}_{\mathbf K_{p})}]+
\widehat{\mathcal C}_{s},\nonumber
\end{align}
where $\widehat{\mathcal C}_{s,o}$, 
$\widehat{\mathcal C}_{s}$, 
when restricted to $\mathcal R^{p-1}$,
are functions of the variables $\Phi^{[p-2]}$. 
It follows that for $p\geq1$,
\begin{eqnarray}
&&\nabla^S_{S'}\Phi^{\mathbf K_{p-1}'S'}_{\mathbf K_{p+1}S} =
\nabla^S_{S'}\nabla^{S'}_{(K_1}\nabla^{(K_{1}'}_{K_{2}}\cdots
\nabla^{K_{p-1}')}_{K_p}\Phi_{K_{p+1}S)}+\nonumber\\
&&\qquad\qquad\qquad
\frac{p^2-p-2+2\delta_{1p}}{p(p+2)}
[\Phi_{S(K_{p+1}},\Phi^{\mathbf K_{p-1}'S}_{\mathbf K_{p})}] +
\widehat{\mathcal C},
\label{E:ContrCovarPhiOrdp}
\end{eqnarray}
where again $\widehat{\mathcal C}$,
when restricted to $\mathcal R^{p-2}$,
are functions of the variables $\Phi^{[p-2]}$.
We use the identity
\[
\nabla^A_{S'}\nabla^{BS'}G = 
\nabla^{(A}_{S'}\nabla^{B)S'}G + 
\frac{1}{2}\e^{AB}\nabla^2G
\]
and equations \eqref{E:wavePhiOrdp}, 
\eqref{E:SkewCovarDer} to compute
\begin{eqnarray}
&&\nabla^S_{S'}\nabla^{S'}_{(K_1}
\nabla^{(K_{1}'}_{K_{2}}\cdots
\nabla^{K_{p-1}')}_{K_p}\Phi_{K_{p+1}S)}=\nonumber\\
&&\qquad\qquad
-\frac{p+1}{p+2}[\Phi_{S(K_{p+1}},
\Phi^{\mathbf K_{p-1}'S}_{\mathbf K_{p})}]-
\frac{p+3}{2(p+2)}\nabla^2\Pvar{K}{}{p+1}{p-1}=
\label{E:ContrCovarOrdpPhiFinal}\\
&&\qquad\qquad=-(\frac{(p+1)^2}{p}-2\delta_{1p})[\Phi_{S(K_{p+1}},
\Phi^{\mathbf K_{p-1}'S}_{\mathbf K_{p})}]-
\nonumber\\
&&\qquad\qquad\qquad\qquad\;\,\frac{(p-1)(p+3)}{p+2}
[\overline\Phi_{S'}^{(K_{p-1}'},
\Phi_{\mathbf K_{p+1}}^{\mathbf K_{p-2}')S'}]+
\widehat{\mathcal D},
\nonumber
\end{eqnarray}
where $\widehat{\mathcal D}$,
when restricted to $\mathcal R^{p-1}$,
depends on the variables $\Phi^{[p-2]}$.

Finally, by \eqref{E:RedSymmComp}, \eqref{E:ATermFinal}, 
\eqref{E:BTermsFinal}, \eqref{E:ContrCovarPhiOrdp}, 
\eqref{E:ContrCovarOrdpPhiFinal}, 
we see that \eqref{E:derPhisymmvar} holds.
\end{bew}


\begin{thebibliography}{99}

\bibitem{sAjP1}
Anco, S. G., Pohjanpelto, J.,
{Classification of conservation laws of Maxwell's equations}.
To appear.

\bibitem{sAjP2}
Anco, S. G., Pohjanpelto, J., 
{Generalized symmetries of Maxwell's equations}.
In preparation. 

\bibitem{Anderson96}
Anderson, I. M., Torre, C. G., 
{Classification of local generalized symmetries 
for the vacuum Einstein equations}. 
Comm. Math. Phys. \textbf{176}, (1996) 479--539.

\bibitem{Glassey79}
Glassey, R. T., Strauss, W. A.,
{Decay of classical Yang-Mills fields}.
Comm. Math. Phys. \textbf{65}, (1979) 1--13.

\bibitem{Helgason78}
Helgason, S., \textit{Differential Geometry, Lie Groups, 
and Symmetric Spaces} (Academic Press, New York 1978).

\bibitem{Kalnins86}
Kalnins, E. G., Miller Jr., W., Williams, G. C.,
{Matrix operator symmetries of the Dirac equation
and separation of variables}.
J. Math. Phys. \textbf{27}, (1986) 1893--1900.

\bibitem{Kalnins92}
Kalnins, E. G., McLenaghan, R. G., Williams, G. C.,
{Symmetry operators for Maxwell's 
equations on curved space-time}.
Proc. R. Soc. Lond. A \textbf{439}, (1992) 103--113.

\bibitem{Kumei75}
Kumei, S.,
{Invariance transformations, invariance group transformations
and invariance groups of the sine-Gordon equations}.
 J. Math. Phys. \textbf{16}, (1975) 2461--2468.

\bibitem{Mikhailov91}
Mikhailov, A. V., Shabat, A. B., Sokolov, V. V.,
{The symmetry approach to classification 
 of integrable equations}.
\textit{What Is Integrability} (V. E. Zakharov ed.)
(Springer-Verlag, Berlin 1991) 115--184.

\bibitem{Miller77}
Miller Jr., W., 
\textit{Symmetry and Separation of Variables}
(Addison-Wesley, Reading, Mass. 1977).

\bibitem{Olver86}
Olver, P. J.,
\textit{Applications of Lie Groups to Differential Equations}
(Springer, New York 1986).

\bibitem{Penrose84}
Penrose, R., Rindler, W.,
\textit{Spinors and Space-time:{Volume 1,} Two-spinor Calculus 
and Relativistic Fields} (Cambridge University Press, Cambridge 1984). 

\bibitem{Schwartz82}
Schwartz, F.,
{Symmetries of SU(2) invariant Yang-Mills theories}.
Lett. Math. Phys. \textbf{6}, (1982) 355--359.

\bibitem{Torre95}
Torre, C. G.,
{Natural symmetries of the Yang-Mills equations}.
J. Math. Phys. \textbf{36}, (1995) 2113--2130.

\bibitem{Tsujishita82}
Tsujishita, T.,
{ On variation bicomplexes associated 
to differential equations}
Osaka J. Math. \textbf{19}, (1982) 311--363.

\bibitem{Vinogradov89}
Vinogradov, A. M.,
{Symmetries and conservation laws of partial
differential equations: Basic notions and results}.
Acta Appl. Math. \textbf{15}, (1989) 3--21.

\bibitem{Ward77}
Ward, R. S.,
{On self-dual gauge fields}.
Phys. Lett. \textbf{61A}, (1977) 81--82.

\bibitem{Ward85}
Ward, R. S.,
{Integrable and solvable systems, 
and relations among them}.
Phil. Trans. R. Soc. Lond. A
\textbf{315}, (1985) 451--457.

\bibitem{Ward90}
Ward, R. S., Wells, Jr., R. O.,
\textit{Twistor Geometry and Field Theory}
(Cambridge University Press, Cambridge 1990).

\end{thebibliography}
\end{document}